\documentclass[12pt]{article}
\usepackage{a4}
\usepackage{epsfig}
\def\Journal#1#2#3#4{{#1} {\bf #2}, #3 (#4)}

\def\NPB{{\em Nucl. Phys.} B}

\def\PRL{\em Phys. Rev. Lett.}
\def\PRD{{\em Phys. Rev.} D}
\def\PRP{{\em Phys. Reports}}

\def\COMP{{\em Comput. Phys. Commun.} }
\begin{document}
\newcommand{\newc}{\newcommand}
\newc{\R}{$R$}
\newc{\charginom}{M_{\tilde \chi}^{+}}
\newc{\mue}{\mu_{\tilde{e}_{iL}}}
\newc{\mud}{\mu_{\tilde{d}_{jL}}}
\newc{\barr}{\begin{eqnarray}}
\newc{\earr}{\end{eqnarray}}
\newc{\beq}{\begin{equation}}
\newc{\eeq}{\end{equation}}
\newc{\ra}{\rightarrow}
\newc{\lam}{\lambda}
\newc{\eps}{\epsilon}
\newc{\gev}{\,GeV}
\newc{\tev}{\,TeV}
\newc{\eq}[1]{(\ref{eq:#1})}
\newc{\eqs}[2]{(\ref{eq:#1},\ref{eq:#2})}
\newc{\etal}{{\it et al.}\ }
\newc{\eg}{{\it e.g.}\ }
\newc{\ie}{{\it i.e.}\ }
\newc{\nonum}{\nonumber}
\newc{\lab}[1]{\label{eq:#1}}
\newc{\dpr}[2]{({#1}\cdot{#2})}
\newc{\gsim}{\stackrel{>}{\sim}}
\newc{\lsim}{\stackrel{<}{\sim}}
\begin{titlepage}
\begin{flushright}
{ETHZ-IPP  PR-98-08} \\
{December 3, 1998}\\
\end{flushright}
\begin{center}
{\bf \LARGE SUSY discovery strategies at the LHC} \\
\end{center}

\smallskip \smallskip \bigskip
\begin{center}
{\Large Michael Dittmar}
\end{center}
\bigskip
\begin{center}
Institute for Particle Physics (IPP), ETH Z\"{u}rich, \\
CH-8093 Z\"{u}rich, Switzerland
\end{center}

\bigskip
\begin{abstract}
\noindent 
Current ideas for SUPERSYMMETRY searches at the LHC are reviewed.
We analyse the discovery prospects for various supersymmetric particles
and describe recent ideas on the possibilities of detailed 
SUSY studies at the LHC. We also combine today's 
experimental knowledge with some speculations about what might 
be known by the year 2005 to provide a realistic picture for 
the LHC start. 
\vspace{2cm}
\end{abstract}

\begin{center}
{\it Lectures given at \\
Summer School on \\
Hidden Symmetries 
and Higgs Phenomena, \\
ZUOZ 16-22 August 1998} \\
\end{center}
\end{titlepage}

\section{Introduction}
The search for new physics phenomena is often defined as the main motivation
for new experiments at higher center of mass energies.
This is especially true for the LHC project, with its two general purpose 
experiments ATLAS and CMS. The prime motivation of the LHC physics 
program is to discover the ``mechanism of electroweak symmetry breaking'' 
usually associated with a scalar particle, the Higgs boson. 
Theoretical ideas suggest that this hypothetical particle with a 
mass of less than roughly 1 TeV could explain the observed mass spectrum 
of bosons and fermions. The simplicity and mathematical elegance of
this model leads however to its own problems, the so called hierarchy problem 
or fine tuning problem of the Standard Model. 

These problems originate 
from theoretical ideas which extrapolate today's knowledge at mass scales of 
a few 100 GeV to energy scales of about $10^{15}$ GeV and more.  
A purely theoretical approach to this extrapolation has led theorists 
to SUPERSYMMETRY which could solve these conceptual problems by the 
introduction of supersymmetric partners to every known 
boson and fermion and at least an additional Higgs super-multiplet. 
Despite the largely unconstrained masses of these new partners,
the potential to discover such new objects has become a central question
for many design issues of future high energy particle physics experiments.  

Following today's theoretical fashions, the main goals of ATLAS and CMS
are often defined as the search for the Higgs boson and for 
any kind of experimental manifestation of supersymmetry. 

A growing fraction of the preparation time for a modern collider 
experiment deals with the simulation of case studies like 
the search sensitivity for new physics. Such studies are 
not only required in order to motivate the required effort, time and money
but should also provide some guidance on ``what is possible'' and 
how a real detector should look like. 

Among the many possible case studies, essentially all simulation studies 
concentrate on the SM Higgs search and on 
SUSY particle searches. The investigated signatures provide thus not only 
``dream-land'' possibilities for a future collider but, as will 
become clear in the following,     
cover a wide range of detector requirements which help to shape the
final and real experiment. 

The following review about SUSY signatures at the LHC 
is structured as follows. We start our discussion with an overview  
of experimental and theoretical constraints for the 
LHC SUSY searches. This is followed by a detailed discussion 
of current ideas about various SUSY signatures at the LHC.
We then speculate about possible studies which can be 
performed once SUSY events have been discovered at the LHC.
Finally we try to combine today's and tomorrows experimental 
constraints to define a realistic SUSY search strategy for the LHC. 
  
\section {The experimental and theoretical frame}

\subsection {The experimental LHC frame}
Table 1 combines the required experimental observables 
with new physics possibilities. 
\begin{table}[htb]
\begin{center}
\begin{tabular}{|c|c|c|}
\hline
Type of measurements & indicates & required for \\
\hline
isolated high $p_{t}$ $e^{\pm}, \mu^{\pm}$ & $W^{(*)}$, $Z^{(*)}$ decays &  
Higgs search \\ 
& &top physics, ``all'' searches \\
\hline
isolated high $p_{t}$ $\gamma$'s & electro-magnetic process & Higgs search \\
\hline
$\tau$ and $b$-quark tagging & ``rare'' processes & 
special Higgs like searches \\
\hline
large missing    & $\nu$ like events & Higgs, Supersymmetry, \\
$p_{t}, E_{t}$   & $W, Z$ decays     & exotic ``exotica''    \\
\hline
jets                   & quarks and gluons & QCD, understanding of \\
                       &                   & backgrounds/efficiencies \\
\hline
\end{tabular}
\caption{New physics and some required detector capabilities.}
\label{table1}
\end{center}
\end{table}
Obviously, real experiments like the LEP or Tevatron 
experiments have to be a compromise between these different requirements.
Nevertheless, the existing experiments have proven to work 
according to or better than specified in their technical proposals.
Especially astonishing results have been achieved 
with silicon-micro vertex detectors, which allow to identify b-flavoured jets 
with high efficiencies and 
excellent purity. In addition, quite accurate 
calorimetry allows to measure the missing
transverse energy in complicated events. Such indirect 
identification of energetic neutrino like objects is now 
routinely used by essentially all large collider experiments.   

The design objectives of the future LHC experiments, ATLAS~\cite{atlastp} and 
CMS~\cite{cmstp},  
follow the above desired detector capabilities with
emphasis on high precision measurement of electrons, muons and 
photons and large angular coverage for jets. 
According to their technical proposals,     
both collaborations 
expect to identify isolated electrons and muons, 
with $p_{t} > 10$ GeV and small backgrounds up to a pseudorapidity 
($\eta = -ln \tan (\Theta/2)$) of $|\eta| \leq 2.5$
and efficiencies of $\approx$ 90\%. Furthermore, both experiments 
expect to achieve b-jet tagging with up to 50\% efficiency and  
light flavour jet rejection factors of up to 100.   
These expectations are used for essentially all simulations of 
LHC measurements. For justifications of these efficiencies
we refer to the various ATLAS and CMS 
technical design reports and internal technical 
notes~\cite{atlascms}. 
  
The above performance figures  
are the essential assumptions for the
simulation results described in the following sections.  
As the expected experimental performance might differ 
to some extend from the future real performance, the critical reader
should include simplicity and robustness as criteria to judge 
today's LHC simulation ``results''.

\subsection {The theoretical MSSM frame}
Among the many possible extensions of the Standard Model the 
Minimal Supersymmetric Standard Model (MSSM) is usually considered 
to be the most serious theoretical frame. The attractive features 
of this approach are:
\begin{itemize}
\item it is quite close to the existing Standard Model; 
\item it explains the so called hierarchy problem of the Standard Model;
\item it allows to calculate and   
\item it predicts many new particles and thus ``Nobel Prizes'' for the 
masses.
\end{itemize}
These attractive features of the MSSM are nicely described 
in a Physics Report from 1984 by H. P. Nilles~\cite{nilles}. 
We repeat here some of his 
arguments given in the introduction: \newline

{\it ``Since its discovery some ten years ago, supersymmetry has 
fascinated many physicists. This has happened despite the absence of 
even the slightest phenomenological indication that it might be relevant 
for nature. .... Let us suppose that the standard model is valid up 
to a grand unification scale or even the Planck scale $10^{19}$ GeV.
The weak interaction scale of 100 GeV is very tiny compared to these 
two scales. If these scales were input parameters of the theory 
the (mass)$^2$ of the scalar particles in the Higgs sector have to 
be chosen with an accuracy of $10^{-34}$ compared to the Planck Mass. 
Theories where such adjustments of incredible accuracy have to 
be made are sometimes called unnatural.... 
Supersymmetry might render the standard model natural... 
To render the standard model supersymmetric a price has to be paid. For 
every boson (fermion) in the standard model, a supersymmetric partner 
fermion (boson) has to be introduced and to construct phenomenological
acceptable models an additional Higgs super-multiplet is needed.''}

Figures 1 and 2~\cite{hollik98} compare the consistency of the various 
electroweak measurements with the SM and the MSSM. 
\begin{figure}[htb]
\begin{center}
\epsfig{file=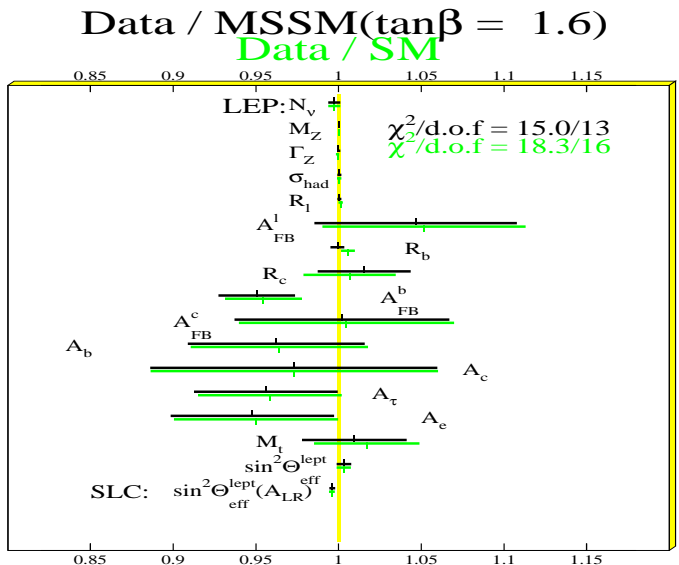,
height=9 cm,width=11cm}
\end{center}
\caption[fig1]{Comparison of $Z^{0}$ precision 
measurements with the Standard Model and the MSSM
with $\tan \beta = 1.6$ and very heavy SUSY particles~\cite{hollik98}.
}
\end{figure}
\begin{figure}[htb]
\begin{center}
\includegraphics*[scale=0.45,bb=0 150 600 650 ]
{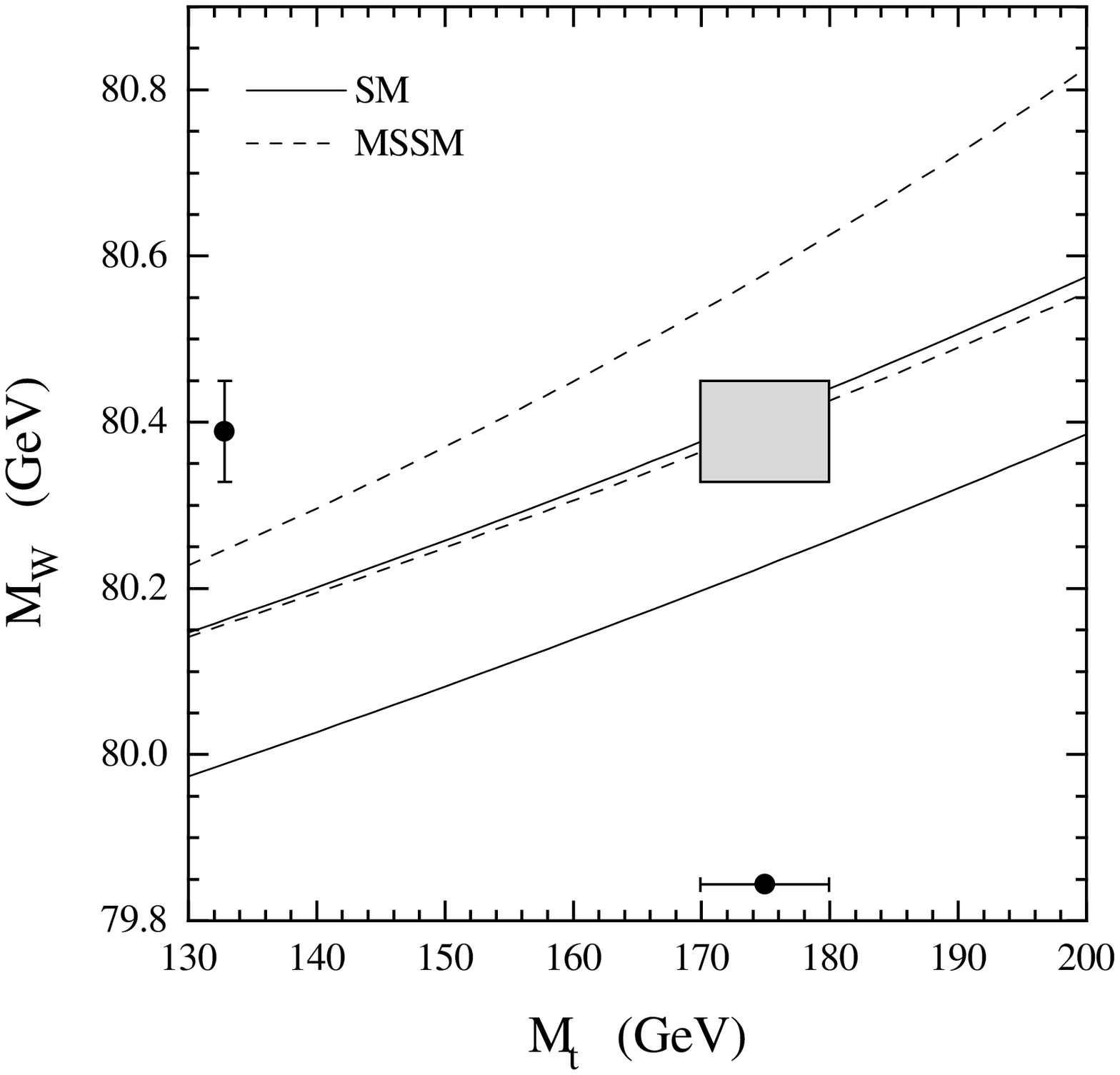}
\end{center}
\caption[fig2]{Expected relation between 
$M_{W}$ and $M_{top}$ in the 
Standard Model and  
the MSSM; the bounds are from the non observation of Higgs or SUSY 
particles at LEPII~\cite{hollik98}.
The 1998 experimental area, with 
$M_{W}=80.39 \pm 0.06$ GeV and $M_{top}=174 \pm 5$ GeV
is also indicated.
}
\end{figure}

The largest difference, shown in Figure 2, appears in the relation 
between the $W^{\pm}$ mass and the top mass. Unfortunately today's data,  
$M_{W}=80.39 \pm 0.06$ GeV and $M_{top}=174 \pm 5$ GeV,   
favour an area which is perfectly consistent with both models.
One might thus conclude that the expected small improvements of 
the electroweak measurements will not allow to distinguish 
between the SM and the MSSM. 

As mentioned above, SUSY predicts a doubling of the fundamental 
fermions and bosons and requires at least 5 Higgs 
bosons\footnote{For discussions about 
LHC signatures for the MSSM Higgs particles
and the sensitivity of the LHC experiments we refer
to~\cite{mssmh}.}.

Beside the lightest, possibly invisible SUSY particle, one knows from  
the absence of such new particles that their masses have to be heavier
than $\approx$ 100 GeV. 
Starting from the MSSM, the so called minimal model, theoretical 
counting results in more than hundred free parameters.
So many free parameters do not offer a good
guidance for experimentalists, who prefer to use additional  
assumptions to constrain the parameter space.
The simplest approach is the so called mSUGRA 
(minimal super-gravity) model with 
only five parameters ($m_{0}, m_{1/2}, \tan \beta, A^{0}$ and $\mu$).

This SUSY model is used for most sensitivity estimates of future 
colliders and the obtained results for the LHC will be discussed below.
The main reason for this model choice is the existence of 
very advanced Monte Carlo programs~\cite{isajet}, \cite{pythia} 
and \cite{spythia},  
required for detailed simulation studies.
This pragmatic choice of one approach to investigate 
the potential of a future experiment appears to be more than sufficient, 
as essentially all required detector features can be tested. 
 
However, this approach should not be considered 
as a too strong guidance principle if one wants to discover 
SUPERSYMMETRY with real experiments. Two recent examples show that   
the absence of any MSUGRA indications enlarges the acceptance for more 
radical SUSY models. 

The first example is the famous lonely CDF event, 
which has large missing transverse energy,
two high $p_{t}$ isolated photons and 2 isolated high $p_{t}$ 
electron candidates~\cite{cdfevent}.
The presence of high $p_{t}$ photons does not match MSUGRA
expectations but might fit 
into so called gauge mediated symmetry breaking models, GMSB~\cite{gmsbth}. 
This event has certainly motivated many additional, 
so far negative searches. 

The second example is related to the 1997 HERA excitement. 
The observed excess of a handful of events was consistent with a 
lepton-quark resonance with a mass of roughly 200 GeV and 
signal predictions from R-parity violation SUSY 
models~\cite{rparity}. While this excess was not confirmed 
with larger statistics, R-parity violation models became 
certainly much more attractive.

These modified searches indicate 
the discovery potential of searches which are 
not guided by today's fashion.
Having reminded the reader of potential shortcomings 
between a SUSY Nature and the studied mSUGRA model, we now turn 
to future LHC (and Tevatron) search strategies for 
SUSY particles within mSUGRA. 

\subsubsection{mSUGRA predictions}

Essentially all signatures related to the MSSM and 
in particular to mSUGRA searches are
based on the consequences of R-parity conservation. R-parity is a 
multiplicative quantum number like ordinary parity. The R-parity of 
the known SM
particles is 1 and -1 for the SUSY partners.
As a consequence, SUSY particles have to be produced in pairs and 
unstable SUSY particles decay, either directly or via 
some cascades, to SM particles and the lightest supersymmetric
particle, the LSP. The LSP, using cosmological arguments, 
is required to be neutral.
As massive LSP's should have been abundantly produced 
after the big bang, the LSP is currently considered 
to be ``the cold dark matter'' candidate. 
This LSP, usually assumed to be the lightest neutralino 
$\tilde{\chi}^{0}_{1}$, has neutrino like interaction cross sections
and can not be observed in collider experiments. 
Events with a large amount of 
missing energy and momentum are thus the prime SUSY  
signature in collider experiments.

A possible example is the pair production of sleptons with 
their subsequent decays, $pp \rightarrow \tilde{\ell}^{+}\tilde{\ell}^{-}$
and $ \tilde{\ell} \rightarrow \ell \tilde{\chi}^{0}_{1}$
which would appear as events with a pair of isolated electrons or muons 
with high $p_{t}$ and large missing transverse energy.  

Within the mSUGRA model, the masses of 
SUSY particles are strongly related to the so called universal fermion 
and scalar masses $m_{1/2}$ and $m_{0}$. The masses of the spin 1/2 
SUSY particles are directly related to $m_{1/2}$. 
One expects approximately the following mass hierarchy:
\begin{itemize}
\item
$\tilde{\chi}^{0}_{1} \approx 1/2 m_{1/2} $
\item
$\tilde{\chi}^{0}_{2}\approx \tilde{\chi}^{\pm}_{1} \approx m_{1/2} $
\item
$\tilde{g}$ (the gluino) $\approx 3 m_{1/2} $
\end{itemize}
The masses of the spin 0 
SUSY particles are related to $m_{0}$ and $m_{1/2}$ and 
allow for some mass splitting between the ``left'' and ``right'' handed
scalar partners of the degenerated left and right handed fermions. 
One finds the following simplified mass relations:
\begin{itemize}
\item
$m(\tilde{q})$(u,d,s,c and b) $ \approx \sqrt{m_{0}^{2} + 6 m_{1/2}^{2}}$
\item
$m(\tilde{\nu}) \approx m(\tilde{\ell^{\pm}})$ (left)
$ \approx \sqrt{m_{0}^{2} + 0.52 m_{1/2}^{2}}$
\item
$m(\tilde{\ell^{\pm}})$ (right) $\approx \sqrt{m_{0}^{2} + 0.15 m_{1/2}^{2}}$
\end{itemize}
The masses of the left and right handed stop quarks ($\tilde{t}_{\ell, r}$) 
might show, depending on other MSUGRA parameters,  
a large splitting. As a result, the right handed stop quark might even be 
the lightest of all squarks.   
 
Following the above mass relations and using the known SUSY couplings, 
possible SUSY decays and the related signatures can be defined.
Already with the simplest mSUGRA frame one finds   
a variety of decay chains.

For example the $\tilde{\chi}^{0}_{2}$ could decay 
to $\tilde{\chi}^{0}_{2}\rightarrow \tilde{\chi}^{0}_{1}+ X$
with $X$ being:
\begin{itemize}
\item
$X= \gamma^{*}, Z^{*} \rightarrow \ell^{+}\ell^{-}$
\item
$X= h^{0} \rightarrow b \bar{b}$
\item
$X= Z  \rightarrow f \bar{f}$
\end{itemize}

Other possible $\tilde{\chi}^{0}_{2}$ decay chains are 
$\tilde{\chi}^{0}_{2}\rightarrow \tilde{\chi}^{\pm(*)}_{1}+ \ell^{\pm} \nu$ 
and 
$\tilde{\chi}^{\pm(*)}_{1} \rightarrow \tilde{\chi}^{0}_{1} \ell^{\pm} \nu$
or 
$\tilde{\chi}^{0}_{2}\rightarrow \tilde{\ell}^{\pm} \ell^{\mp}$.

Allowing for higher and higher masses, even more decay channels
might open up. It is thus not possible to define all search strategies
a priori. Furthermore, possible 
unconstrained mixing angles between neutralinos, lead to 
model dependent decay chains and
search strategies for squarks and gluinos as 
will be discussed below. 

Today's negative SUSY searches~\cite{treille} 
provide the following approximate lower mass 
limits:
\begin{itemize}
\item
$m(\tilde{\chi}^{\pm}_{1}) > 90$ GeV (LEPII) 
\item $m (\tilde{g}) $(gluino) $>$ 160-220 GeV depending slightly
on the assumed relation between squark and gluino masses (TEVATRON).
\end{itemize}

One might argue that the negative results of the chargino search at LEPII
imply that future gluino searches at the upgraded TEVATRON 
should not start for masses below $\approx$ 270 GeV.
However, the continuing TEVATRON searches indicate that many 
searchers do not follow too strictly specific mass relations of the
mSUGRA model. Current experimental results are usually shown  
as a function of the searched for SUSY masses.
\begin{figure}[htb]
\begin{center}
\includegraphics*[scale=0.7,bb=40 320 600 800 ]
{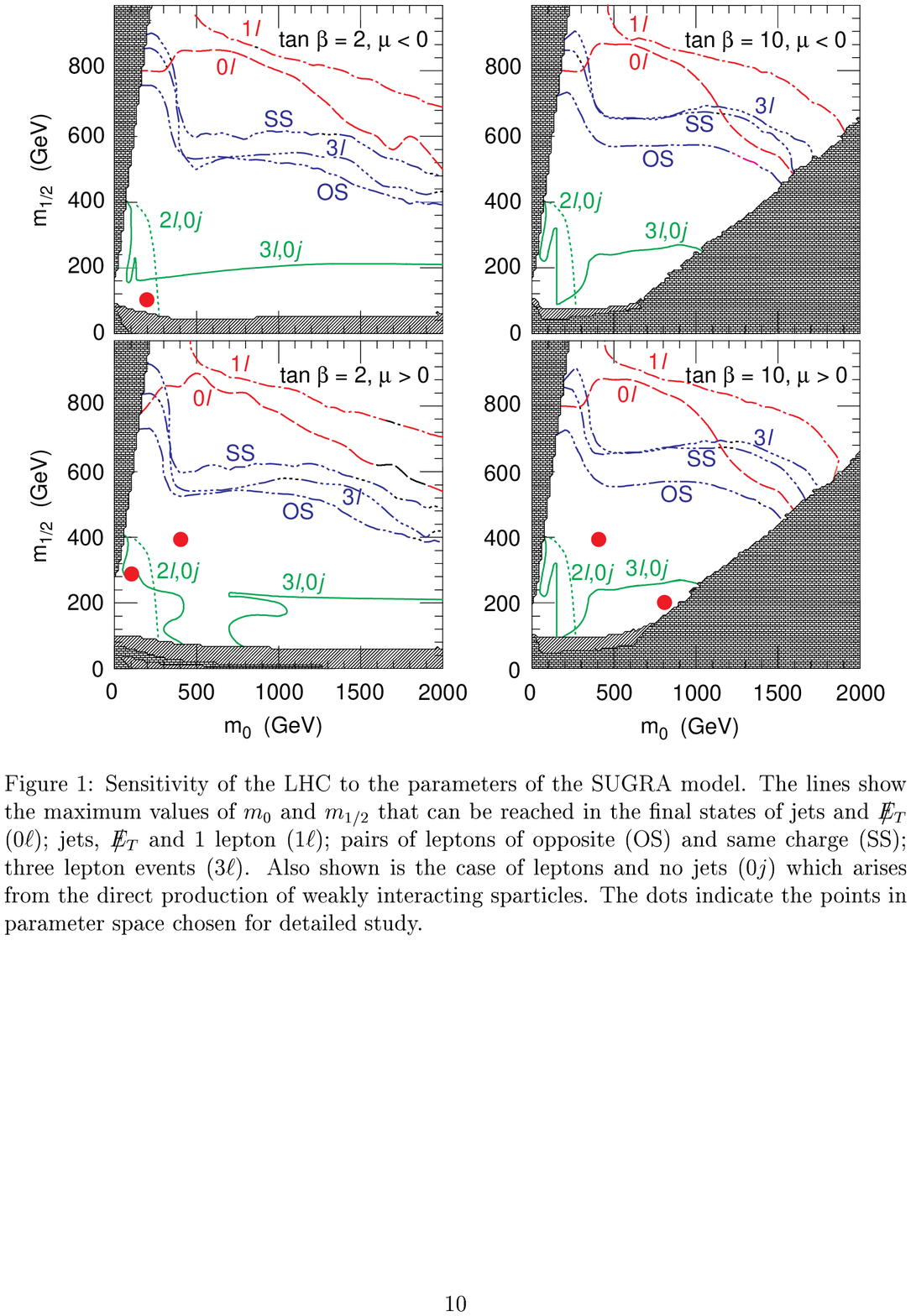}
\caption[fig3]
{Expected mSUGRA sensitivity of various signatures 
in the $m_{0} - m_{1/2}$ plane at the LHC,
assuming an integrated luminosity of 
10 fb$^{-1}$~\cite{baer96}. The different curves indicate 
the expected sensitivity for SUSY events with n leptons ($\ell$) 
and for events with lepton pairs with same charge (SS) and 
opposite charge (OS).}   
\end{center}
\end{figure}

In contrast, sensitivity estimates for future collider experiments are 
usually given in the $m_{0}$--$m_{1/2}$ parameter space. 
Despite the model dependence, such estimates allow to 
compare the possible significance of the different 
analysed signatures. 
Having various proposed methods, the resulting 
sensitivity figures appear to be quite confusing and require 
some time for appreciation.   
A typical example is shown in Figure 3a-d~\cite{baer96}, 
where the different 
curves indicate the LHC sensitivity for different signatures 
and different SUSY particles. It is usually assumed that the maximum 
information about SUSY can be extracted in regions covered 
by many signatures. The meaning of the various curves and their potential
significance should become clear from the following sections.  
\subsection{Searches, Significance and Systematics}

Peaks in the invariant mass spectrum of assumed decay 
products are an unambiguous  
signature for new unknown particles. Narrow mass 
peaks can in principle be discovered without the help of
any theoretical Monte Carlo programs 
as backgrounds can reliably be estimated from sidebands.
Furthermore, quite accurate significance estimates can be made 
even if very small signal (S) to background (B) ratios
are expected and if not all background sources are known.
The reason becomes quite obvious from the following example with  
an expected Signal of 1000 events above a 
background of 10000. Such a deviation from known sources
could be claimed with a significance of about 10 standard deviations
($N(\sigma) = S/\sqrt{B} = 1000/\sqrt{10000}$). Assuming a relatively 
smooth flat background over many non signal bins, 
background extrapolations to the signal region 
can reach systematic accuracies of less than a percent.
Under such ideal conditions, even large background uncertainties are 
acceptable as the background could go up by a factor of about 4, still giving 
a 5 standard deviation signal! 
Nevertheless at least signal Monte Carlos are needed to 
determine cross sections and perhaps 
other quantities like spin and parity
once a mass peak signal is observed. 
Furthermore, even searches for mass peaks require usually 
some selection criteria to isolate possible signals from
obvious backgrounds. Searchers should however  
remember that advantages of optimised efficiencies, obtained with
complicated selection methods, 
are easily destroyed by uncontrolled systematic errors.
Other disadvantages of too much optimisation are     
model dependent phase space restrictions and the introduction 
of statistical fluctuations which 
increase in proportion with the number of cuts and mass bins.

In addition, it is not always an advantage to reduce signal and backgrounds 
to relatively small numbers when the significance has to be calculated 
from Poisson statistics! For example a simple $\sqrt{B}$ estimate 
for 9 expected background events requires an observation of 
at least 24 events, e.g. an excess of 15 events above 9 background events,
to claim a 5 $\sigma$ excess above background. However, for small event 
numbers one finds that the $\sigma = \sqrt{B}$ approximation is not 
good enough 
a 5 $\sigma$ excess requirement, equivalent to 
a background fluctuation probability of  
less than $6\times 10^{-7}$. 
Using Poisson statistics, the required 5 $\sigma$ excess 
corresponds to an observation  
of more than 27 events! Despite this reduced significance 
(roughly $1\sigma$), systematic errors
start to become important. For small background numbers 
the sideband method is limited by statistics and direct and clean 
background estimates from data might increase/decrease backgrounds 
and might be larger than Monte Carlo background  
estimates. The method to determine backgrounds,  
either from data or from Monte Carlo might 
thus hide or enhance a real signal and 
artificial good or bad limits can be 
obtained\footnote{The critical reader should investigate 
how often search limits appear to be ``lucky'', 
e.g. the number of observed 
data events is smaller than the number of expected background events,
and how often they are ``unlucky''.} 

The possibility to observe fluctuations 
due to many mass bins appears nicely 
in a CMS simulation~\cite{ecaltdr} of the  
two photon mass distribution for a SM Higgs with a 
mass of 130 GeV and backgrounds.
\begin{figure}[htb]
\begin{center}
\mbox{
\epsfig{file=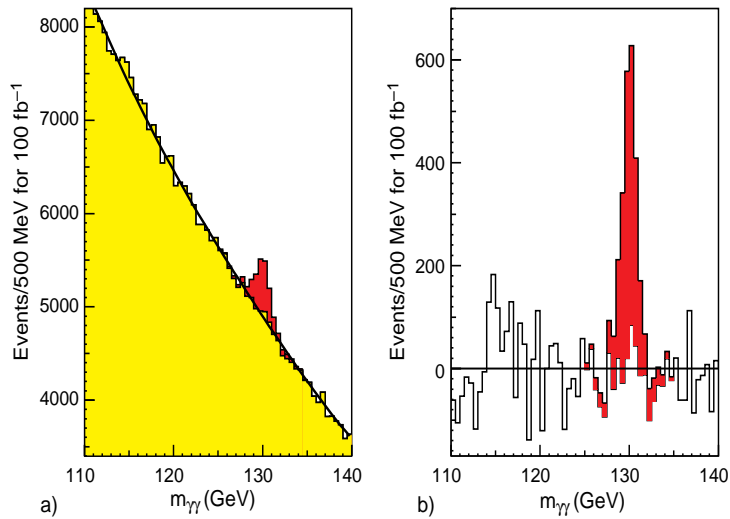,
height=7 cm,width=10cm}
}
\end{center}
\caption[fig4]
{CMS search simulation~\cite{ecaltdr} for 
$H \rightarrow \gamma \gamma$ before and after background 
subtraction.}  
\end{figure}
Figure 4 shows 
a clear narrow peak is seen at 130 GeV. The observed signal,  
assuming a simple straight line to estimate the background, 
corresponds to about 10 standard deviations.
A more careful analysis of the mass distribution shows 
why at least five standard deviations are required to 
establish the existence of a new particle.  
Ignoring for example the simulated Higgs signal at 130 GeV
one might try to look for an excess of events at an arbitrary 
mass. The largest excess of events appears at a mass of about 115 GeV. 
Taking the background from the sidebands one finds 
a statistical fluctuation with a significance of about 
three standard deviations. Thus, many possible mass bins combined with 
various event selection criteria are a remaining danger for mass peak hunters.

Despite the simplicity to discover new physics with mass peaks, 
most searches for new physics phenomena 
require an excess of events in special kinematic 
regions or tails of distributions. Some difficulties of such searches are 
indicated in Figure 5.
\begin{figure}[htb]
\begin{center}
\mbox{
\epsfig{file=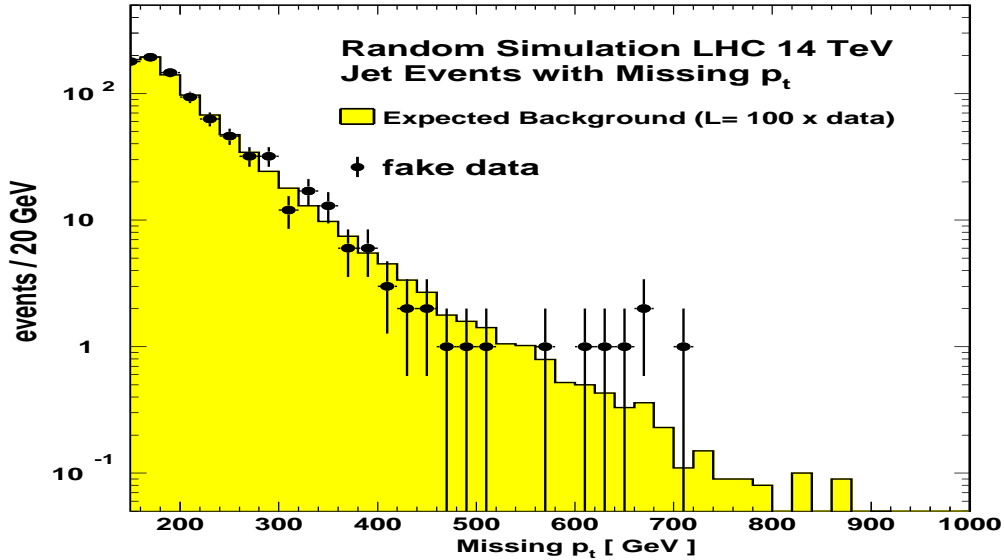,
 bbllx=100pt,bblly=150pt,bburx=500pt,bbury=680pt,
height=8 cm,width=10cm}
}
\end{center}
\caption[fig5]
{Simulation of a background fluctuation for events with 
missing transverse energy.}  
\end{figure}

The figure shows a random simulation of missing transverse energy 
events from $pp \rightarrow Z X \rightarrow \nu \bar{\nu} X$  
and small statistics which is compared to  
a background simulation with large statistics of the same process.   
Depending on the new physics signature, the small excess 
of tail events might coincide with a signal, expected for a certain
range of missing transverse momentum $p_{t}$. For a missing $p_{t}$ 
between 600-720 GeV one could quote an excess of almost 3 sigma, e.g.
6 events are seen while a background of only 2 events are predicted.
``Good arguments'' might increase 
the significance for new physics further.  
For example one might argue that the Monte Carlo overestimates the 
backgrounds, as the sideband region between 400--500 GeV 
shows about 50\% more events than found with the ``pseudo data''. 
Some additional features of the 6 events might
further be used to increase the significance.
Alternatively, the possibility of new physics can easily 
be excluded from the same distribution, for example one could 
argue that the number of 7.63 predicted events with missing $p_{t}$ 
above 500 GeV is in perfect agreement with the observed 8 events.   
The above example justifies the statement 
{\bf ``never search in tails''}. 
Unfortunately, most new physics scenarios, like SUPERSYMMETRY, would 
appear as rare events and in tails of distributions. 

Thus, ingenuity is required to separate new physics from tails of known 
processes. Such searches require not only to have enough statistical 
significance but {\bf a method to determine backgrounds}. 
The difficulty to 
establish a signal becomes clear from the following two examples.
{\bf Case a} is for a comfortable Signal to Background ratio of 1:1
while {\bf case b} is for a ratio of 1:10. The required minimal statistics 
is easy to estimate. A 5 sigma excess needs a statistics of
roughly 25 Signal events on top of a background 
of 25$\pm$5 for {\bf case a} while {\bf case b} 
needs about 250 signal events above a background of 2500$\pm$50.
The statistical excess however is not enough as 
the expected background has some systematic errors 
like uncertainties from the efficiency, the luminosity and the 
theoretical background model. Assuming that all these uncertainties 
are known with an accuracy of $\pm 5$\%, the 
significance of {\bf case a} is essentially unchanged while 
the significance of {\bf case b} is reduced to about 2 standard deviations.

It is worth noting that some studies claim to be {\bf ``conservative''} 
by multiplying backgrounds by arbitrary factors (method 1)  
or by using the error estimate from $\sqrt{S+B}$ (method 2).

Using {\bf case b} and method 1 one sees no 
dramatic change of the estimated sensitivity. One finds that only the 
minimal luminosity requirement has to be increased.
At the same time, the signal to background ratio became 1:20 and a $\pm$5\%
systematic error would result in almost meaningless results!

The estimated sensitivity, 
using method 2, does not change for a very bad signal to background 
ratio. However, one finds that 
method 2 reduces a clear signal, like 10 observed events with 
one expected background event, to a modest 3 standard deviation signal. 

We thus disagree with the 
claim that the above methods are conservative and reliable. 
In contrast, a correct approach 
to a possible significance figure of an ``average'' future experiment
should give the statistical sensitivity for new physics, estimated with 
$\sigma \approx \sqrt{B}$, and 
has to describe how backgrounds can be estimated 
and how well they need to be known. Most sensitivity estimates 
do not provide answers to the latter requirements.
Attention should thus be paid to the estimated signal 
to background ratio, which allows to estimate the 
required systematic accuracies.

\section {Anatomy of a Signal, Searching for Sleptons at the LHC}

Hadron colliders are certainly not a good source of sleptons. 
Nevertheless, we start our analysis of the various SUSY search strategies 
with an anatomy of the simplest possible SUSY signal.
Our discussion starts with the cross section and the 
expected decay modes. This is followed by a qualitative 
description of a possible discovery signature at the LHC
which is then compared to a detailed simulation of a search for sleptons
at the LHC. 

The pair production of sleptons at the LHC can easily be related to the 
production of Drell-Yan dilepton pairs with high mass.
The expected total slepton pair production cross section 
as a function of the slepton mass is shown in Figure 6~\cite{baer94}.
 
\begin{figure}[htb]
\begin{center}
\includegraphics*[scale=0.7,bb=40 350 600 800 ]
{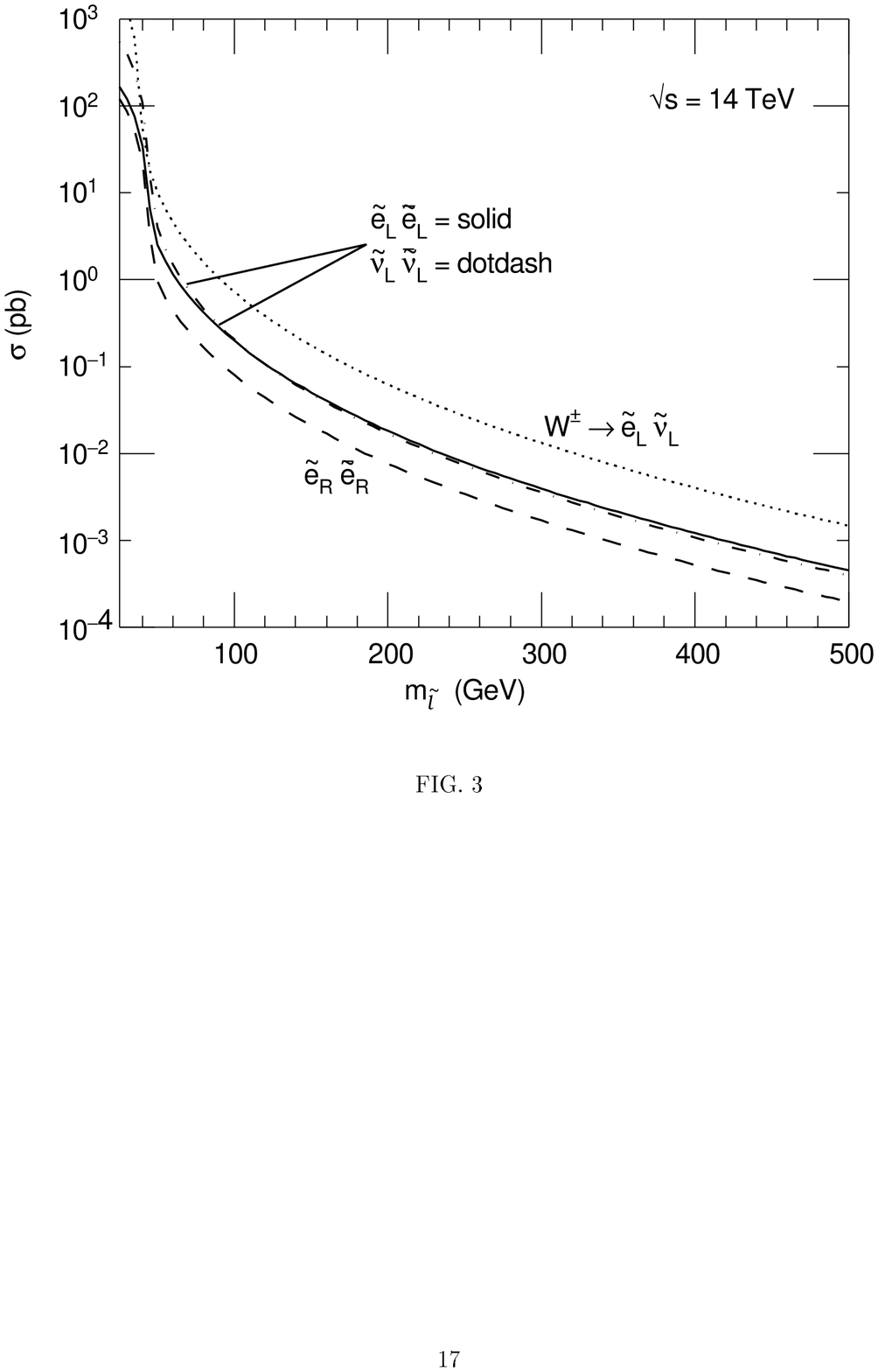}
\caption[fig6]
{Slepton Mass dependence of the total pair production cross section 
at the LHC for various slepton combinations~\cite{baer94}.}
\end{center}
\end{figure}

Figure 7 shows the expected mass (with m$>$200 GeV) 
distribution of the virtual $\gamma^{*}, Z^{*}$ 
system leading to a lepton or slepton pair. 
The cross section for scalar charged sleptons has a simple relation 
to the production of the corresponding right and left handed lepton 
pair production 
$\sigma(\tilde{\ell} \tilde{\ell})=1/4 \beta^{3} \sigma(\ell\ell)$.
\begin{figure}[htb]
\begin{center}
\includegraphics*[scale=0.6,bb=40 150 600 700 ]
{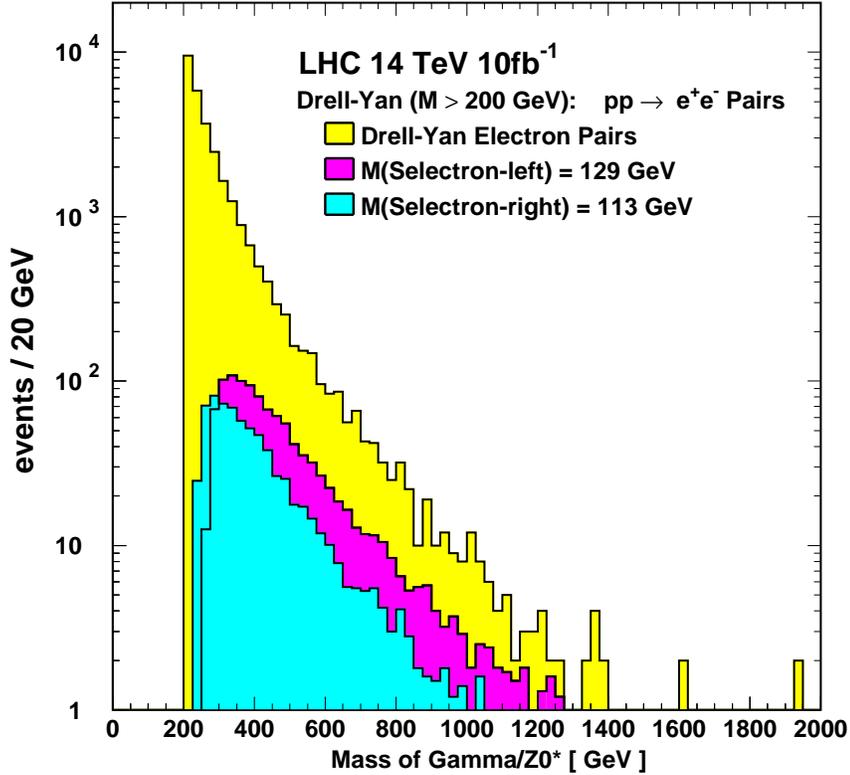}
\caption[fig7]
{Mass distribution of Drell-Yan electron pairs with a mass above 200 GeV 
and for left or right handed selectron pairs at the 
LHC with selectron masses of 129 GeV and 113 GeV respectively
as obtained with PYTHIA~\cite{pythia} and SPYTHIA~\cite{spythia}.}    
\end{center}
\end{figure}
For our example the masses of the left handed and right handed slepton 
were fixed to 129 GeV and 113 GeV respectively. The 
expected mass spectra show the $\beta^{3}$ cross section suppression 
close to threshold.  The larger rate for Drell-Yan pairs produced  
from the left handed virtual $\gamma^{*}, Z^{*}$ system results, despite the
larger mass, into a bigger cross section for left handed sleptons.
This simple relation between slepton mass and cross section
allows precise cross section predictions for slepton pairs,
once the corresponding 
mass spectrum of Drell-Yan lepton pairs has been measured.

As a next step one has to consider the possible slepton decay modes. 
While the right handed slepton can decay only to 
the lightest neutralino and the corresponding lepton 
$\tilde{\ell}^{\pm} \rightarrow \tilde{\chi}^{0}_{1} \ell^{\pm}$, 
several somehow model dependent decay modes, 
are possible for left handed sleptons:
\begin{center}
\begin{itemize}
\item
$\tilde{\ell}^{\pm} \rightarrow \tilde{\chi}^{0}_{1} \ell^{\pm}$
~~~or~~~ 
$\tilde{\ell}^{\pm} \rightarrow \tilde{\chi}^{0}_{2} \ell^{\pm}$
~~~or~~~ 
$\tilde{\ell}^{\pm} \rightarrow \tilde{\chi}^{\pm}_{1} \nu$
\item
$\tilde{\nu} \rightarrow \tilde{\chi}^{0}_{1} \nu~~$
~~~~or~~~ 
$\tilde{\nu} \rightarrow \tilde{\chi}^{\pm}_{1} \ell^{\mp}$
\end{itemize}
\end{center}

The best signature for  
slepton pair production appears to be the two body 
decay $\tilde{\ell}^{\pm} \rightarrow \tilde{\chi}^{0}_{1} \ell^{\pm}$.
The resulting signature are events with a pair of two isolated 
same flavour leptons with opposite charge and some missing transverse 
momentum. To distinguish such signal events from various 
possible SM backgrounds a long list of kinematic selection
criteria have to be applied. To identify good selection criteria
at the LHC it is useful to start with simplified     
kinematics in the center-of-mass frame.  
The observable leptons, originating from the decays of massive sleptons, 
should show: (1) a characteristic momentum spectrum; (2) should not 
balance their 
momenta and (3) should not be 
back to back. Furthermore, the measurable mass of the event should 
be much smaller than the original center-of-mass energy and the missing mass
should be much larger than zero. 

A possible selection requires thus the possibility to measure 
isolated leptons with good accuracy and to determine indirectly 
the missing energy and momentum from all detectable particles.
An accurate missing energy determination requires 
an almost 4$\pi$ acceptance for all visible particles.
Unfortunately, a realistic experiment has to live with several 
detection gaps especially the ones around the beam pipe. 
Consequently, missing momentum measurements along the beam direction 
are of limited use. In addition, the event kinematics 
at a Hadron Collider are unfortunately very different from the 
center-of-mass frame. As a result, signal and background events have a
large and unknown momentum component along the beam direction. 
However, variables which exploit the missing transverse
energy and momentum remain very useful.

Furthermore, in contrast to an $e^{+}e^{-}$ collider with a fixed 
dilepton mass, hadron collider searches must consider 
the effects that slepton pairs are not produced at a fixed $\sqrt{s}$
and show a wide spread of the longitudinal momentum.
As a result, good selection criteria exploit the differences between signal
and background in the plane transverse to the beam.
Such variables are (1) the transverse momenta of each lepton, 
(2) the opening angle between the two leptons in the plane transverse to the 
beam and (3) the missing transverse momentum. 
The specific choice of cuts depends strongly on the 
studied mass region and the relevant backgrounds.
 
The largest ``irreducible'' background for slepton pair production 
are events with leptonic $W^{\pm}$ decays from  
$W$-pair production $pp \rightarrow WW X$ with 
($\sigma \times BR(WW \rightarrow e^{+} \nu e^{-} \bar{\nu}$) of about 0.8 pb.
Another potentially very large background comes from 
$t \bar{t}$ production with a 
$\sigma \times BR (t\bar{t} \rightarrow 
WW b \bar{b} \rightarrow e^{+} \nu e^{-} \bar{\nu} X)$ 
with a cross section of about 7 pb. This background can be strongly reduced 
by applying a jet veto.
Other potential backgrounds are miss-measured Drell-Yan lepton pairs and 
electrons and muons from leptonic $\tau$ decays produced in the reaction   
$pp \rightarrow \tau \tau$. 
Additional backgrounds, usually assumed to be negligible,
might come from events of the type $W^{\pm} X$ and $Z^{0} X$  
with leptonic decays and isolated high $p_{t}$ hadrons which are 
misidentified as electrons or muons.   
In addition, other unknown sources of new physics, like 
$pp \rightarrow \tilde{\chi}^{+} \tilde{\chi}^{-}$ might also result
in events with two isolated leptons and missing transverse momentum.
 
The qualitative ideas discussed above, can now be compared with 
a quantitative simulation of a slepton search with 
the CMS experiment~\cite{denegri96059}.

The analysis selects first events which contain 
a pair of opposite charged electrons or muons and no additional jets.
It is assumed that isolated electrons or muons with 
a minimum transverse momentum of 20 GeV and $|\eta| < 2.5$
can be identified with high efficiency ($ \epsilon > 90$ \% )
and small backgrounds.
One assumes also that jets with a transverse energy above 30 GeV
and $|\eta| < 4.5$ can be identified and vetoed. 
In addition, events from $pp \rightarrow Z \rightarrow \ell^{+}\ell^{-}$ 
are rejected by demanding that 
the invariant mass of the lepton pair should be inconsistent 
with a $Z^{0}$. Additional mass dependent 
selection criteria, specified in table 2, 
are required to improve the potential signal significance.
\begin{table}[htb]
\begin{center}
\begin{tabular}{|c|c|c|c|c|c|}
\hline
$m (\tilde{\ell}^{\pm})$ & $p_{t}^{lepton} $ & $E_{t}(miss) $ &
$\Delta \phi_{\ell^{+}\ell^{-}}$ & S (100 fb$^{-1}$) & 
B (100 fb$^{-1}$) \\
\hline
100 GeV &  $>$ 20 GeV & $>$ 50 GeV & $>$ 130$^{o}$ & $\approx$ 3200 & 
$\approx$ 10000 \\
\hline
200 GeV &  $>$ 50 GeV & $>$ 100 GeV & $<$ 130$^{o}$ & $\approx$ 230 & 
$\approx$ 170 \\
\hline
300 GeV &  $>$ 60 GeV & $>$ 150 GeV & $<$ 130$^{o}$ & $\approx$  67 & 
$\approx$ 45 \\
\hline
400 GeV &  $>$ 60 GeV & $>$ 150 GeV & $<$ 140$^{o}$ & $\approx$  24 & 
$\approx$ 53 \\
\hline
\end{tabular}
\caption{CMS simulation of the charged slepton search
at LHC~\cite{denegri96059}. The proposed selection criteria 
and signal (S) and background (B) rates are given for a luminosity 
of 100 fb$^{-1}$ and a few slepton masses.}
\label{table2}
\end{center}
\end{table}

The double leptonic decays from $W$ pairs, originating from 
$W^{+}W^{-}$ and $t \bar{t}$, appear to be the dominant backgrounds.
For a slepton mass of about 100 GeV one finds a statistical 
significant signal of $\approx$ 300 events 
above a background of about 1000 events and a luminosity of 10 fb$^{-1}$.
The expected signal and background distributions before the 
$\Delta \phi$ cut are shown in Figure 8a and b. 

\begin{figure}[htb]
\begin{center}
\includegraphics*[scale=0.65,bb=180 120 415 650 ]
{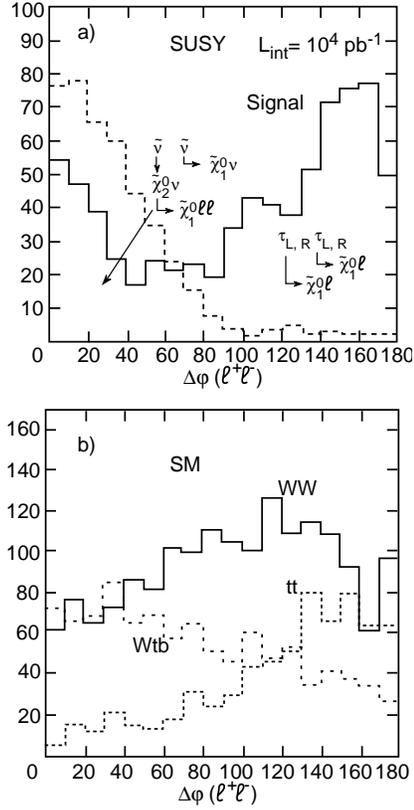}
\caption[fig8]
{Relative azimuthal angle $\phi$ between the two leptons 
for (a) sleptons with masses close to 100 GeV
and other SUSY signals and (b) for the main SM 
backgrounds~\cite{denegri96059}. The proposed cut 
for this slepton mass region is $\phi>130^{o}$.}
\end{center}
\end{figure}

The analysis shows that sleptons with masses between 200-300 GeV
can be selected with signal to background ratios of about 1:1. 
The low signal cross section requires however a large luminosity
of at least 30 fb$^{-1}$. 
For larger masses the slepton cross section becomes very small  
and seems to limit the mass reach to about 400 GeV 
with expected signal rates of 24 events and a total expected background    
of about 50 events for a luminosity of about 100 fb$^{-1}$.
In summary, pair production of charged sleptons at the LHC 
appears to be detectable from an excess of events above 
dominant backgrounds from leptonic decays of $W^{+}W^{-}$ 
and $t\bar{t}$ events. The expected mass reach starts from  
about 100 GeV, roughly the final LEPII reach, and is limited 
to masses of about 400 GeV. 
Particular problems are the 
signal to background ratio for masses well below 200 GeV and the
small signal rate for masses above 300 GeV. 

\clearpage
Other slepton signals, like the one from the reaction 
$pp \rightarrow W^{*} \rightarrow \tilde{\ell} \tilde{\nu}$ 
have been studied and were found to be hopeless~\cite{baer94}.
The investigated signature of
a single high $p_{t}$ lepton with large missing 
$E_{t}$ was found to be at least two orders of magnitude 
smaller than the event rate from single $W$'s as shown in Figure 9. 
The authors concluded further that a possible 
trilepton signal, from $\tilde{\ell} \rightarrow \ell \tilde{\chi}^{0}_{1}$
and cascade decays of the sneutrino 
$\tilde{\nu} \rightarrow \tilde{\chi}^{0}_{2} \nu \rightarrow \ell \ell  
\tilde{\chi}^{0}_{1}$ is much smaller than the signal from 
the simultaneous produced trilepton events of the type
$\tilde{\chi}^{0}_{2}\tilde{\chi}^{\pm}_{1}$ described in the next section. 

\begin{figure}[htb]
\begin{center}
\includegraphics*[scale=0.7,bb=40 350 600 800 ]
{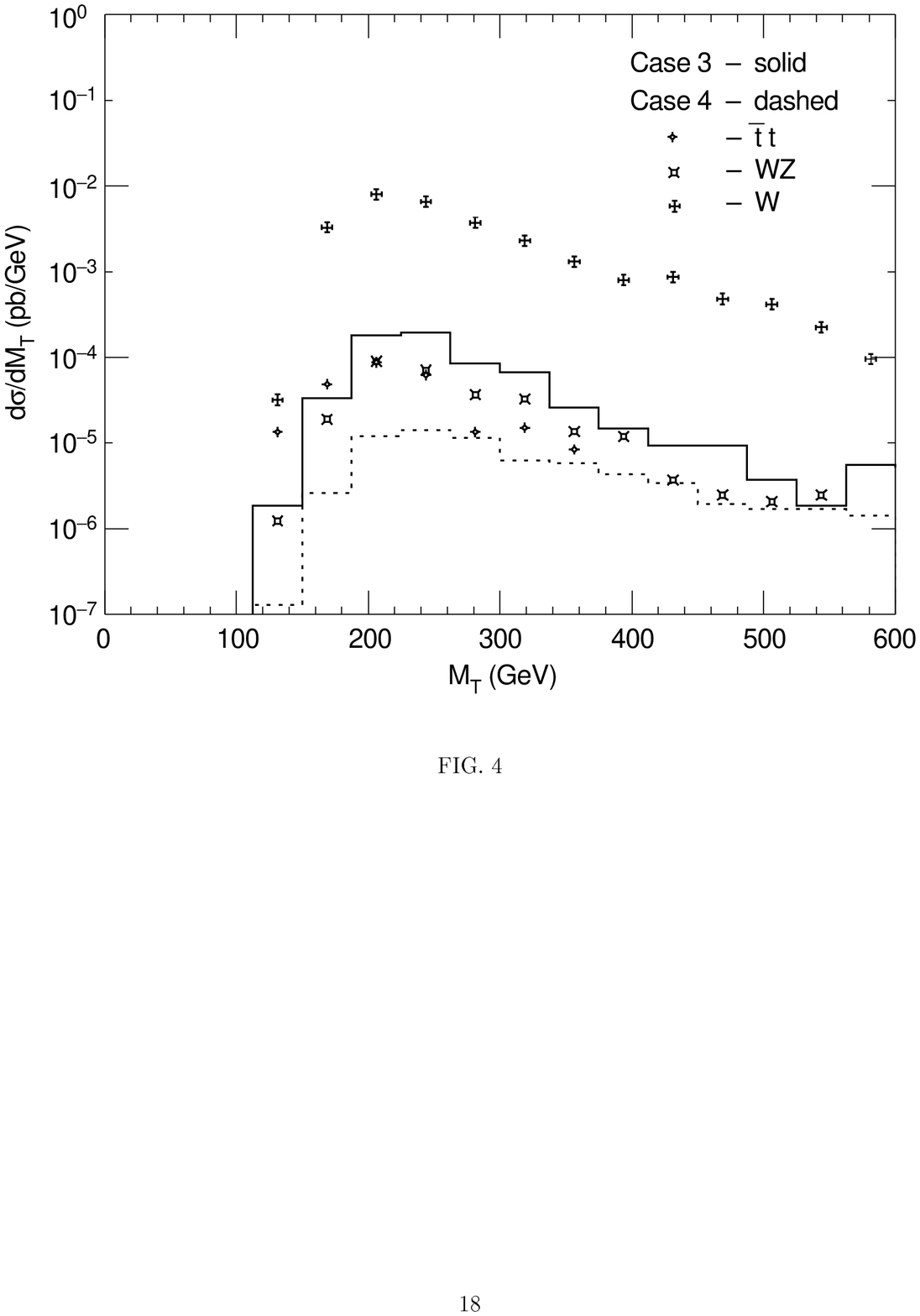}
\caption[fig9]
{Reconstructed transverse mass of single high $p_{t}$ lepton events
for signal events of the type 
$pp \rightarrow W^{*} \rightarrow \tilde{\ell}^{\pm} \tilde{\nu}
\rightarrow \ell^{\pm} \nu \tilde{\chi}_{1}^{0}\tilde{\chi}_{1}^{0}$ 
and various backgrounds. The studied slepton masses were
100 GeV (case 3) and 200 GeV (case 4)~\cite{baer94}.}
\end{center}
\end{figure}

\section {The Trilepton event signature, a signal for 
Chargino-Neutralino Pair Production}

In analogy to the reaction 
$q\bar{q} \rightarrow Z^{0} W^{\pm}$, one might expect the production of 
$q\bar{q} \rightarrow \tilde{\chi}^{0}_{2}\tilde{\chi}^{\pm}_{1}$ events.
Such events might be detected from an analysis of events with 
three isolated high $p_{t}$ leptons and large missing transverse energy. 
The potential of this trilepton signature at hadron colliders, 
like the LHC, has been described in several phenomenological 
studies~\cite{trileptons}. It was found that trilepton events 
with jets should be rejected to
distinguish signal events from SM and SUSY backgrounds.

After the removal of jet events the only remaining relevant 
background comes from leptonic decays of $WZ$ events.
Potential backgrounds from dilepton events like 
$W^{+}W^{-} \rightarrow \ell^{+} \nu \ell^{-} \bar{\nu}$ and 
hadrons misidentified as electrons or muons 
are usually assumed to be negligible.
Depending on the analysed SUSY mass range, the background from 
leptonic decays of $WZ$ events, in contrast to a 
potential signal, will show a $Z^{0}$ mass peak in the dilepton 
spectrum. 

This signature is also used at the Tevatron. 
Estimates for RUN II, with a few fb$^{-1}$, hope for a 
$\tilde{\chi}^{0}_{2}\tilde{\chi}^{\pm}_{1}$ mass sensitivity  
of up to 130 GeV, which might be improved further to about 210 GeV
with RUN III (20-30 fb$^{-1}$)~\cite{tev2000susy}. 
These estimates assume 
a background cross section of less than 0.5 fb.
This number can be compared to recent searches for trilepton events, 
optimised for masses of $\approx$ 80 GeV, from 
CDF~\cite{cdftrileptons}. 
Table 3 shows the current CDF background estimates for various applied cuts 
resulting in a final background cross section of about 10 fb.
\begin{table}[htb]
\begin{center}
\begin{tabular}{|c|c|c|c|}
\hline
Selection     & observed  & SM Background & MSSM MC \\
criteria      & Events    & Expectation & $M(\tilde{\chi}_{1}^{\pm},
M(\tilde{\chi}_{2}^{0}$ =70 GeV \\
\hline
Dilepton data                & 3270488 &  &   \\
\hline
Trilepton data               &     59  &  &  \\
\hline
Lepton Isolation             &     23  &  &   \\
\hline
$\Delta R_{\ell \ell} > 0.4$ &      9  &  &   \\
\hline
$\Delta \phi_{\ell\ell}<170^{o}$ &  8 & 9.6$\pm$1.5 & 6.2 $\pm$0.6  \\
\hline
$J/\Psi, \Upsilon, Z$ removal &     6 & 6.6$\pm$1.1 & 5.5 $\pm$0.5  \\
\hline
missing $E_{t}(miss) > 15 $  &      0 & 1.0$\pm$0.2 & 4.5 $\pm$0.4  \\
\hline
\end{tabular}
\caption{Results from a recent trilepton 
analysis from CDF with a dataset of $\approx$ 
100 pb$^{-1}$~\cite{cdftrileptons}.
The number of observed events appears to be in good agreement 
with various SM 
background sources, which are unfortunately 
given only for the last three cuts.}
\label{table3}
\end{center}
\end{table}

A recent CMS simulation~\cite{cms97007} 
of the trilepton signal at the LHC proceeds as 
follows:

\begin{itemize}
\item
Events should contain three isolated leptons, all with $p_{t} > 15$ GeV
and $|\eta| < 2.5$ and no jets. 
\item
The missing transverse energy should exceed 15 GeV.
\item
The possible same flavour dilepton mass combinations should be 
inconsistent with a $Z^{0}$ decay.  
\end{itemize}

Depending on the studied mass range, additional or harder 
selection criteria are applied. Figure 10 shows the expected 
missing transverse energy distribution for trilepton signal events, 
with different choices of $m_{0}$ and $m_{1/2}$, and for 
background events.
Table 4 gives a few numbers for signal and backgrounds from the CMS
study and different SUSY masses. 
\begin{table}[htb]
\begin{center}
\begin{tabular}{|c|c|c|c|}
\hline
$M_{1/2} \approx M(\tilde{\chi}_{1}^{\pm})$ & $\sigma \times$ BR (trileptons)  
& Signal & SM Background  \\
          &              & (100 fb$^{-1}$) & (100 fb$^{-1}$) \\
\hline
100         & 0.8~~-1.3 pb & 4000--8000    &  900   \\
\hline
150         & 0.04-0.08 pb &    300-600    & 1000   \\
\hline
200         & 0.01-0.02 pb &  80-120       &  700   \\
\hline
300-400     & 0.01-0.02 pb &      50       &  100   \\
\hline
\end{tabular}
\end{center}
\caption{Expected signal and background numbers from a CMS trilepton 
study with different choices of $m_{0}$ and $m_{1/2}$ 
with $\tan \beta = 2$ and negative $\mu$~\cite{cms97007}.}
\label{table4}
\end{table}
\begin{figure}[htb]
\begin{center}
\includegraphics*[scale=0.8,bb=40 100 600 750 ]
{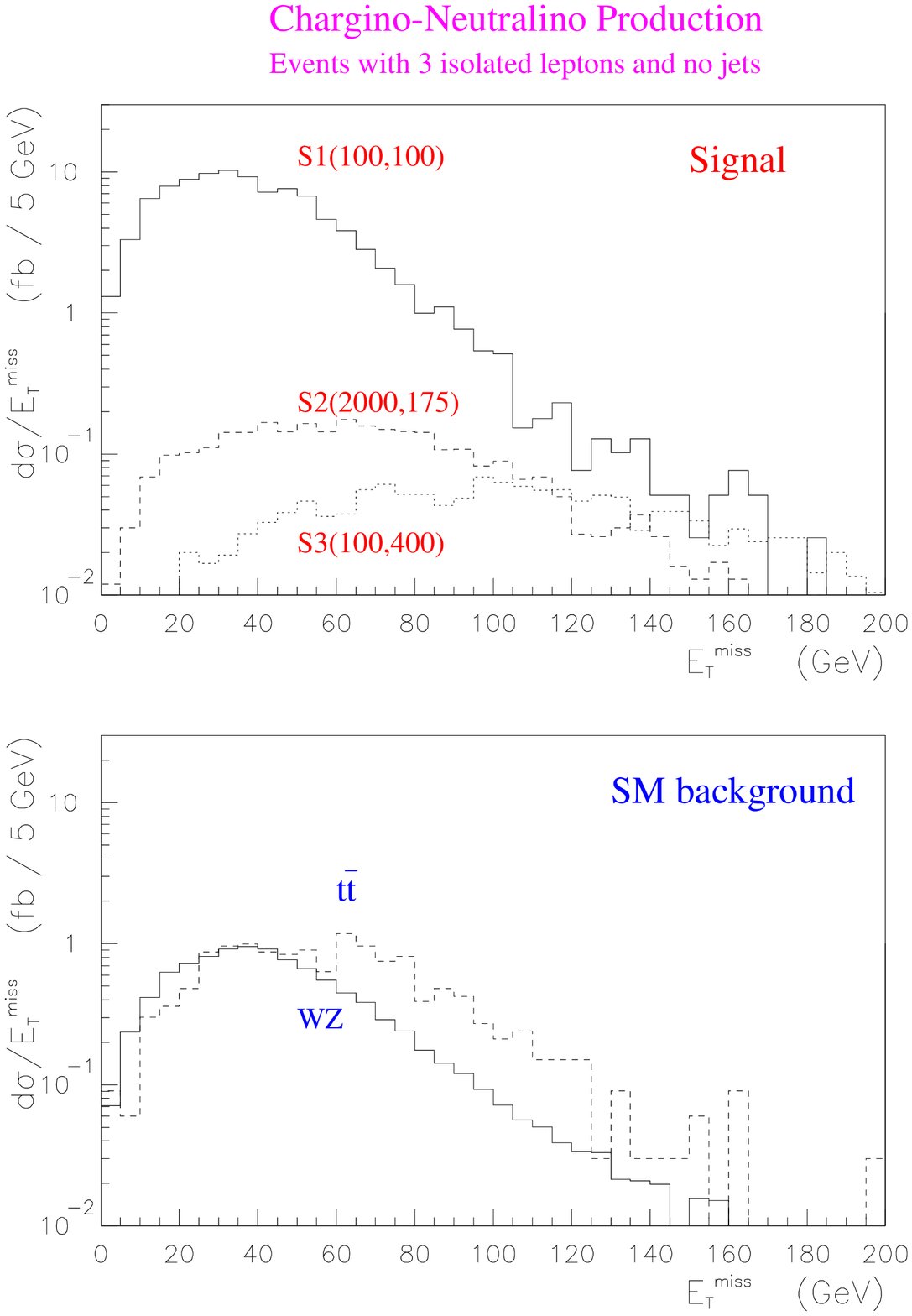}
\caption[fig10]
{Missing transverse energy distribution for 
trilepton events without jets from 
$pp \rightarrow \tilde{\chi}^{0}_{2}\tilde{\chi}^{\pm}_{1}$ signal events  
and background~\cite{cms97007}. The numbers in brackets 
give the used $m_{0}$ and $m_{1/2}$ values.}
\end{center}
\end{figure}

In all cases one finds signal efficiencies of $\approx$ 5\%. 
The best results are obtained for masses close to 100 GeV 
with expected signal rates of $\approx 40$ above a background of 10 events
per 1 fb$^{-1}$ of luminosity. The signal rate drops quickly
for higher masses and much higher luminosities are required 
to establish potential signals up to masses of at most 300-400 GeV. 
Furthermore, for some $m_{0}, m_{1/2}$ mass regions, the estimated 
leptonic branching ratios are very small and result in
signal to background ratios smaller than 0.2.
We conclude that the LHC experiments 
can measure excellent trileptons signals 
in mass and parameter regions where the discovery has most probably been made 
at the upgraded Tevatron RUNII and RUN III by CDF/D0. 
Such high statistics signals will allow 
some detailed SUSY studies as described in section 7.5.
For chargino/neutralino masses above $\approx$ 200 GeV significant 
signals require at least 30 fb$^{-1}$ and a very good understanding of 
possible backgrounds. However, as will become clear from 
the next section, cascade decays of squarks and gluinos 
should provide a much better sensitivity for 
charginos and neutralinos with higher masses.

\clearpage

\section {Squark and Gluino Searches, the Hadron Collider show case}
The discussion in the previous sections covered the potential 
to study non-hadronic interacting SUSY particles with 
relatively small cross section. We now turn the discussion to the 
search for squarks and gluinos with large couplings to quarks and 
gluons.  
The cross section for strongly interacting particles at hadron colliders
like the LHC are quite large.
For example the pair production cross section of 
squarks and gluinos with a mass of $\approx$ 1 TeV has been estimated 
to be as large as 1 pb resulting in 
10$^{4}$ produced SUSY events for one ``low'' luminosity 
LHC year. Such high rates,
combined with the possibility to observe many different 
decay modes, is often considered as a {\it raison d'\^{e}tre} for the LHC.

Depending on the SUSY model parameters, a large variety of 
massive squark and gluino decay channels and signatures might exist.
A complete search analysis for squarks and gluons 
at the LHC should consider the various 
signatures resulting from the following decay channels. 
\begin{itemize}
\item $\tilde{g} \rightarrow \tilde{q} q$ and perhaps 
$\tilde{g} \rightarrow \tilde{t} t$ 
\item $\tilde{q} \rightarrow \tilde{\chi}^{0}_{1} q$ ~~~or~~~
$\tilde{q} \rightarrow \tilde{\chi}^{0}_{2} q$  ~~~or~~~
$\tilde{q} \rightarrow \tilde{\chi}^{\pm}_{1} q$
\item $\tilde{\chi}^{0}_{2} \rightarrow \tilde{\chi}^{0}_{1} \ell^{+}\ell^{-}$ 
 ~~~or~~~ $\tilde{\chi}^{0}_{2} \rightarrow \tilde{\chi}^{0}_{1} Z^{0}$  ~~~or~~~
$\tilde{\chi}^{0}_{2} \rightarrow \tilde{\chi}^{0}_{1} h^{0}$
\item $\tilde{\chi}^{\pm}_{1} \rightarrow 
\tilde{\chi}^{0}_{1} \ell^{\pm}\nu$   ~~~or~~~
$\tilde{\chi}^{\pm}_{1} \rightarrow \tilde{\chi}^{0}_{1} W$.
\end{itemize}  
The various decay channels 
can be separated into at least three distinct event signatures.
\begin{itemize}
\item Multi-jets plus missing transverse energy. These events 
should be spherical in the plane transverse to the beam.
\item Multi-jets plus missing transverse energy plus n(=1,2,3,4) 
isolated high $p_{t}$ leptons. These leptons originate 
from cascade decays of charginos and neutralinos.
\item
Multi-jets plus missing transverse energy plus same charge lepton pairs. 
Such events can be produced in events of the type 
$\tilde{g}\tilde{g} \rightarrow \tilde{u} \bar{u} \tilde{d} \bar{d}$
with subsequent decays of the squarks to 
$\tilde{u} \rightarrow \tilde{\chi}^{+} d$ 
and $\tilde{d} \rightarrow \tilde{\chi}^{+} u$ with 
subsequent leptonic chargino decays  
$\tilde{\chi}^{+} \rightarrow \tilde{\chi}^{0}_{1} \ell^{+} \nu$.  
\end{itemize}
It is easy to imagine that the observation and detailed analysis
of the different types of SUSY events might allow
the discovery of many SUSY particles and should 
help to measure some of the many MSSM parameters.

The above signatures have already been investigated with the data from 
the Tevatron RUN I. The negative searches gave 
mass limits for squarks and gluinos 
as high as $\approx 200$ GeV. The estimated 
5-sigma sensitivity for RUN II and RUN III 
reaches values as high as 350-400 GeV.
More details about the considered signal and backgrounds 
can be found from the TeV2000 studies~\cite{tev2000susy} and 
the ongoing Tevatron workshop.

\begin{figure}[h]
\begin{center}
\includegraphics*[scale=0.7,bb=40 200 600 700 ]
{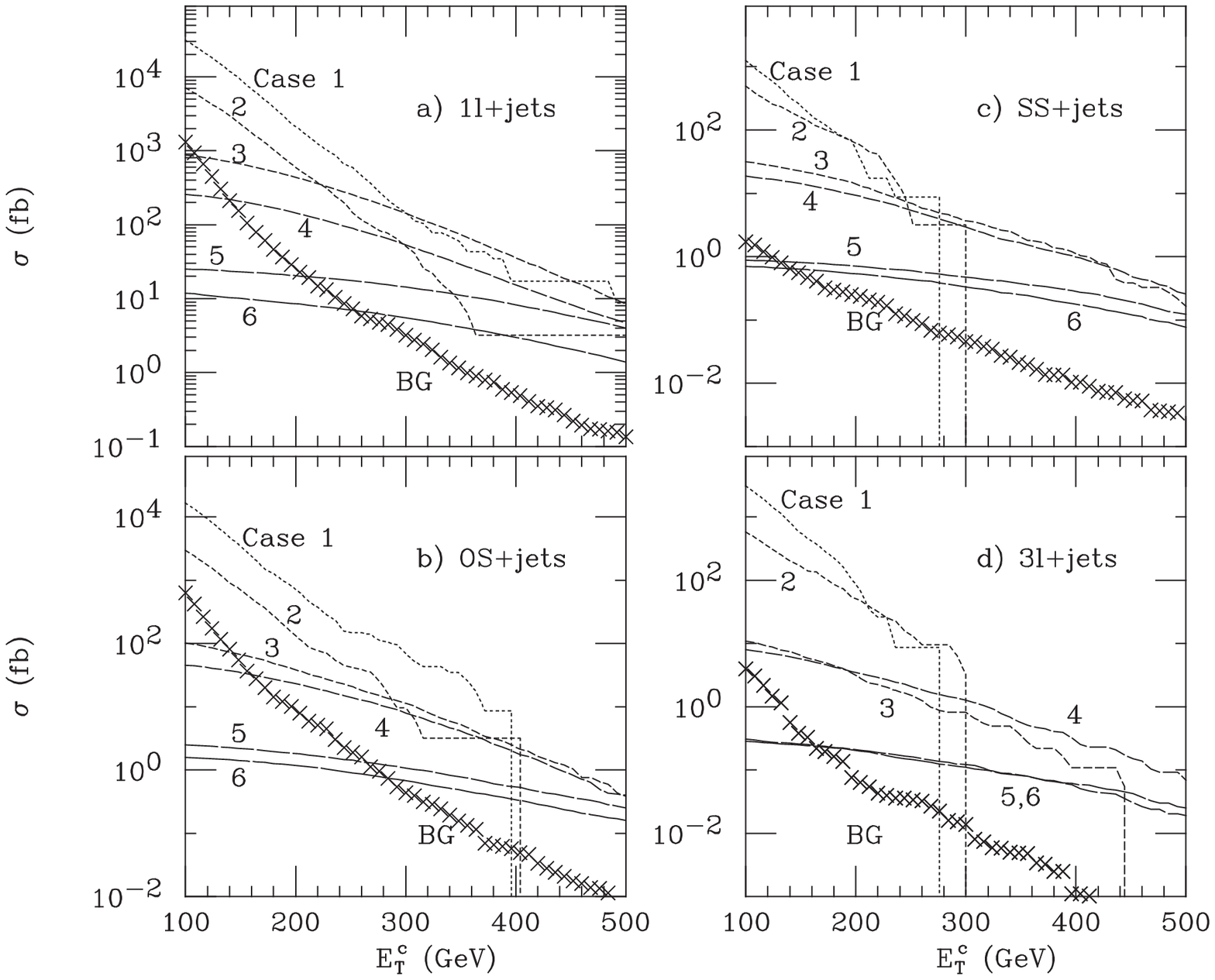}
\caption[fig11]
{Expected $E_{t}^{c}$ distributions for SUSY signal 
and background processes at the LHC and realistic experimental cuts
for $\tan \beta = 2$ and $\mu < 0$~\cite{baer96}.
The different cases are for: 
(1) $m_{\tilde{g}}$= 290 GeV and $m_{\tilde{q}}$= 270 GeV;  
(2) $m_{\tilde{g}}$= 310 GeV and $m_{\tilde{q}}$= 460 GeV;  
(3) $m_{\tilde{g}}$= 770 GeV and $m_{\tilde{q}}$= 720 GeV;
(4) $m_{\tilde{g}}$= 830 GeV and $m_{\tilde{q}}$= 1350 GeV;  
(5) $m_{\tilde{g}}$= 1400 GeV and $m_{\tilde{q}}$= 1300 GeV;
(6) $m_{\tilde{g}}$= 1300 GeV and $m_{\tilde{q}}$= 2200 GeV.}
\end{center}
\end{figure}

\begin{figure}[htb]
\begin{center}
\includegraphics*[scale=0.7,bb=40 200 600 700 ]
{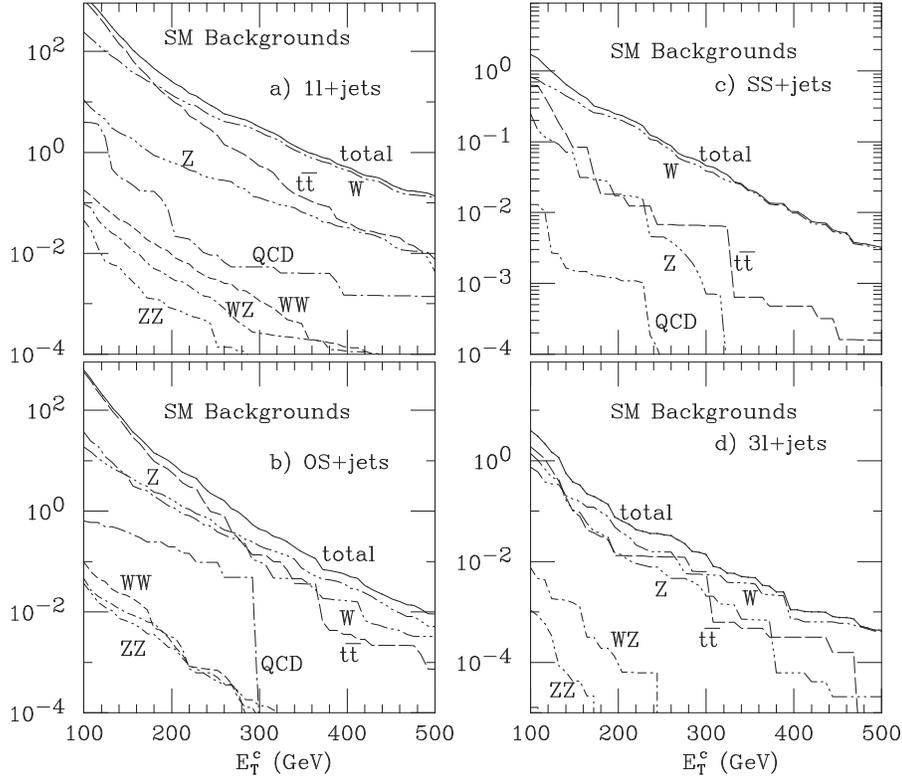}
\caption[fig12]
{Expected $E_{t}^{c}$ distribution for various backgrounds 
after some minimal event selection criteria are applied~\cite{baer96}.}
\end{center}
\end{figure}

A simplified search strategy for squarks and gluinos 
at the LHC would study 
jet events with large visible transverse mass and some missing 
transverse energy. Such events can then be classified according 
to the number of isolated high $p_{t}$ leptons.
Once an excess above 
SM backgrounds is observed for any possible combination of the 
transverse energy spectra, one would try to explain the 
observed types of exotic events and their cross section(s) for
different SUSY $\tilde{g}, \tilde{q}$ masses and decay modes and models.
An interesting approach to such a multi-parameter analysis uses some 
simplified selection variables. For example one could use
the number of observed jets, leptons, their transverse momentum, 
the missing transverse momentum and the visible transverse energy and mass 
to separate signal and 
backgrounds. Such an approach has been used to perform a ``complete''
systematic study of $\tilde{g}$ and $\tilde{q}$ decays~\cite{baer96}
and~\cite{baer95}. 
The proposed variable $E_{t}^{c}$ is the value of the smallest of 
$E_{t}$(miss), $E_{t}$(jet1), $E_{t}$(jet2). The events are further 
separated into the number of isolated leptons. Events with lepton pairs 
are divided into same sign (charge) pairs (SS) and opposite 
charged pairs (OS). 
Signal and background distributions for various squark and gluino 
masses, obtained with such an approach are shown in Figure 11.  

According to this classification, the number of expected 
signal events can be compared with the various SM background processes.
The largest and most difficult backgrounds originate, as can be seen from 
Figure 12, mainly from 
$W+$jet(s), $Z+$jet(s) and $t\bar{t}$ events. 

Using this approach, very encouraging
signal to background ratios, combined with quite large signal cross sections 
are obtainable for a large range of squark and gluino masses.
The simulation results indicate, as shown in Figure 13,
that the LHC experiments with a luminosity of 100 fb$^{-1}$ are sensitive to 
squark and gluinos with masses as high as 2 TeV.
\begin{figure}[htb]
\begin{center}
\includegraphics*[scale=0.7,bb=40 50 600 580 ]
{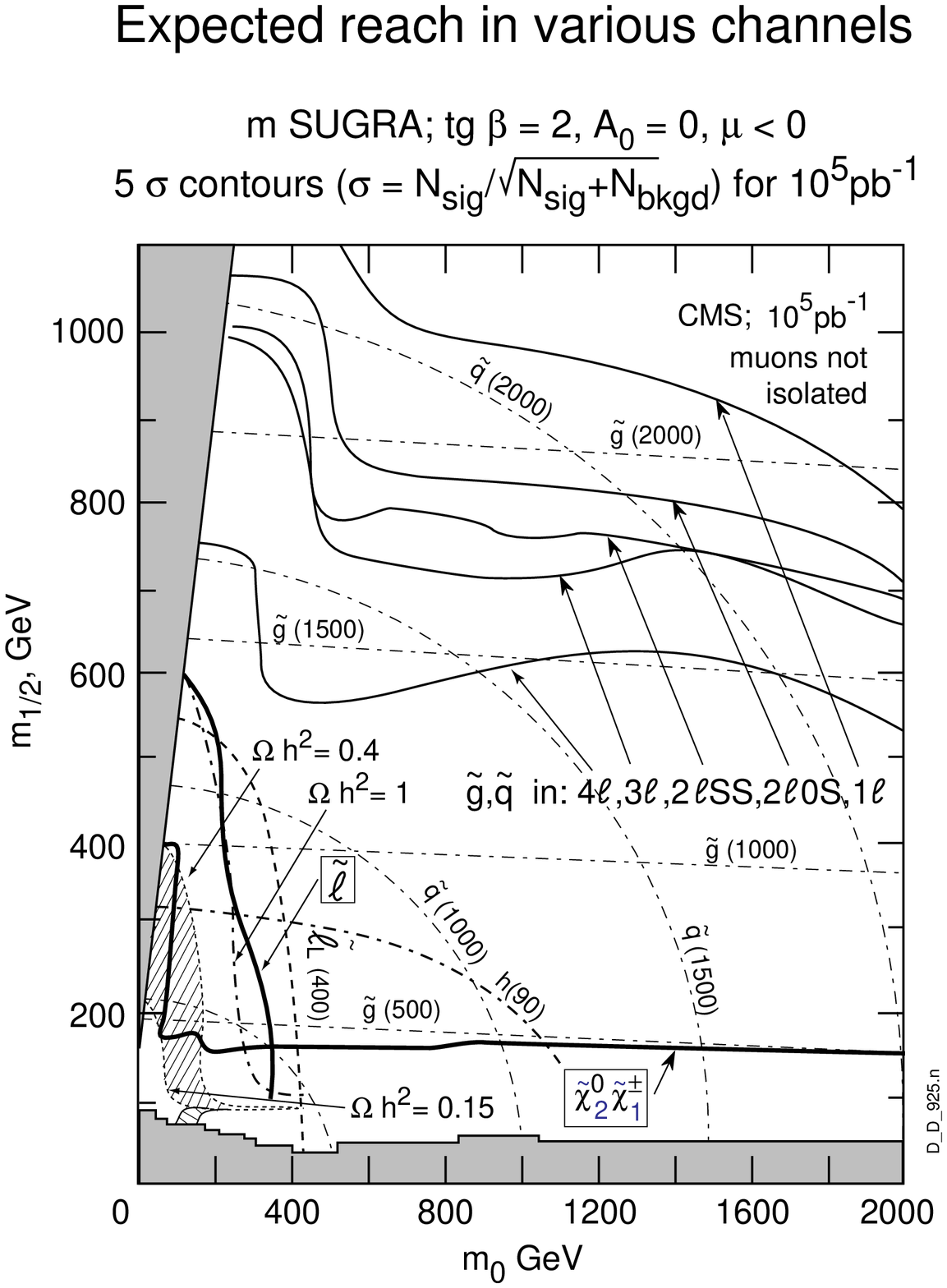}
\caption[fig13]
{Expected ultimate (L=100 fb$^{-1}$) CMS sensitivity 
for squarks and gluinos, sleptons and for 
$\tilde{\chi}^{0}_{2}\tilde{\chi}^{\pm}_{1}$
in the $m_{0}-m_{1/2}$ plane and $\tan \beta = 2$ and $\mu < 0$.
The different full lines show the 
expected 5 sigma signal, estimated from S/$\sqrt{S+B_{SM}}$, 
coverage domain for the various signatures
with isolated high $p_{t}$ 
leptons~\cite{cms98006}. The dashed lines indicate the 
corresponding squark and gluino masses.}
\end{center}
\end{figure}

Figure 13 indicates further, that detailed studies of branching ratios 
are possible up to squark or gluino masses of about 1.5 TeV, where 
significant signals can be observed with many different channels.   
Another consequence of the expected large signal cross sections
is the possibility that the ``first day'' LHC luminosity 
$\approx$ 100 pb$^{-1}$ should be sufficient to discover 
squarks and gluinos up to masses of about 600-700 GeV, 
well beyond even the most optimistic 
Tevatron Run III mass range.

Having this exciting discovery potential for squarks and gluinos 
with many different channels,
one might want to know the ``discovery'' or simply the ``best'' channel.
Such a question is unfortunately not easy to answer. 
All potential signals depend strongly on a good understanding  
of various backgrounds and thus the detector systematics. 
Especially the requirements of high efficiency lepton identification 
and a good missing transverse energy measurement demand for a
``perfect'' working and understood detector. 
This requirement of a good understanding 
of complicated ``monster'' like experiments needs thus some time
and is in contradiction with the ``first day'' discovery potential.
We conclude that the best discovery signature is not yet known,
but should be one which is extremely robust and simple and should  
not depend on too sophisticated detector elements and their resolutions.

\section {After a SUSY Discovery .. what can be measured at the LHC}

Our discussion of the LHC SUSY discovery potential 
has demonstrated the sensitivity of 
the proposed ATLAS and CMS experiments. Being convinced of this 
discovery potential, one certainly wants to know if 
``the discovery'' is consistent with SUPERSYMMETRY and
if some of the many SUSY parameters can be measured.   

To answer this question one should try to find many
SUSY particles 
and measure their decay patterns as accurately as 
possible. The sensitivity 
of direct exclusive SUSY particle production at the LHC has 
demonstrated the various possibilities and 
cross section limitations for weakly produced SUSY particles.
\begin{figure}[htb]
\begin{center}
\includegraphics*[scale=0.5,bb=-80 -200 400 400 height=1. cm,width=8.cm]
{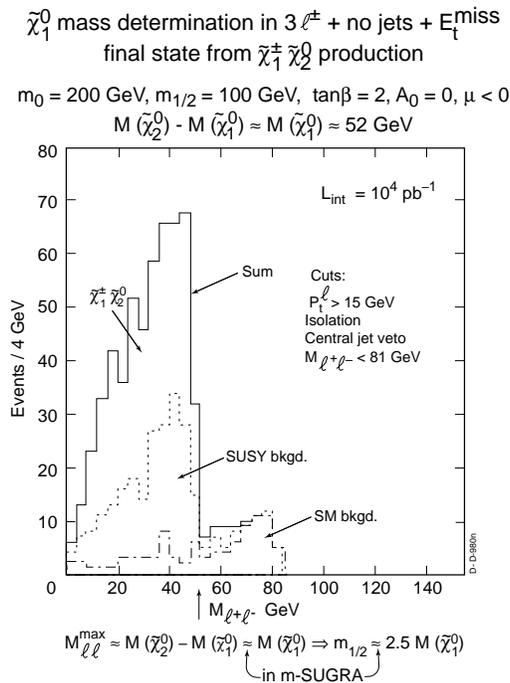}
\caption[fig14]
{Expected CMS dilepton mass distribution with L=10 fb$^{-1}$ and
trilepton events from 
$\tilde{\chi}^{0}_{2}\tilde{\chi}^{\pm}_{1}$~\cite{cms98006}.
The edge in the distribution at about 
50 GeV corresponds to the kinematic 
limit in the decay 
$\tilde{\chi}^{0}_{2} \rightarrow \ell^{+}\ell^{-}\tilde{\chi}^{0}_{1}$
and is thus sensitive to the mass difference 
between $\tilde{\chi}^{0}_{2}$ and $\tilde{\chi}^{0}_{1}$.}
\end{center}
\end{figure}
 
Nevertheless, one finds that the
production and decays of $\tilde{\chi}^{0}_{2}\tilde{\chi}^{\pm}_{1}$
provide good rates for masses below 200 GeV and should allow, 
as indicated in Figure 14, to measure accurately the dilepton mass 
distribution and their relative $p_{t}$ spectra.
The mass distribution and especially the edges 
in the mass distribution are sensitive
to the mass difference between the two neutralinos.
Depending on the used mSUGRA parameters, one finds that the 
$\tilde{\chi}^{0}_{2}$ can have two or three body decays. 
The relative $p_{t}$ spectra of the two leptons
can be used to distinguish the two possibilities.

\begin{figure}[htb]
\begin{center}
\includegraphics*[scale=0.45,bb=100 200 480 700]
{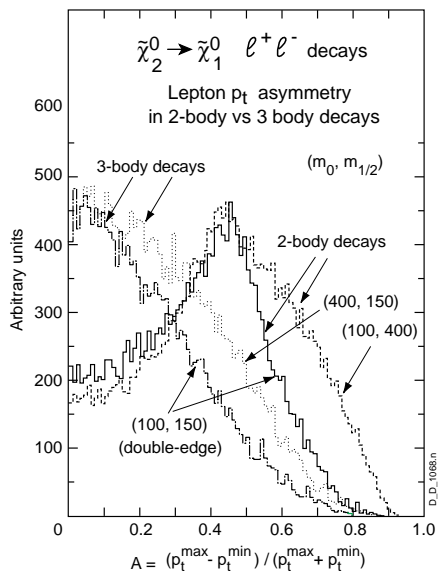}
\caption[fig15]
{Lepton $p_{t}$ asymmetry distributions A in $\tilde{\chi}^{0}_{2}$ decays,
for different choices of $m_{0}$ and $m_{1/2}$ 
and dilepton masses below and above the edge at $\approx$
50 GeV as shown in figure 14~\cite{cms98006}.}
\end{center}
\end{figure}
 
Figure 15 shows the distribution for the variable $A$, defined as 
$A =(p_{t}^{max} -p_{t}^{min}) /(p_{t}^{max} + p_{t}^{min})$ in 
trilepton events and dilepton masses below and above 50 GeV. 
This asymmetry variable originates from 
early investigations of $\tau$ decays~\cite{taudis} where it   
allowed to demonstrate that the leptonic $\tau$ decays 
$\tau \rightarrow \ell \nu \nu$ are three body decays. 
 
In contrast to the rate limitations of weakly produced SUSY 
particles at the LHC, detailed 
studies of the clean squark and gluino events are expected 
to reveal much more information.  
In detail, one finds that the large rate for many 
distinct event channels allows to measure masses and mass 
ratios for several SUSY particles, which are possibly 
being produced in cascade decays of squarks and gluons. 
Many of these ideas have been discussed at a 1996 
CERN Workshop~\cite{cernth96}.
Especially interesting appears to be the idea that 
the $h^{0}$ might be produced and detected in the decay chain 
$\tilde{\chi}^{0}_{2} \rightarrow \tilde{\chi}^{0}_{1} h^{0} $ and 
$h^{0} \rightarrow b \bar{b}$. The simulated mass distribution 
for $b \bar{b}$ jets 
in events with large missing transverse energy is shown 
in Figure 16. Clear Higgs mass peaks above background are found 
for various choices of $\tan \beta$ and $m_{0}, m_{1/2}$.  

\begin{figure}[htb]
\begin{center}
\includegraphics*[scale=1.2,bb=0 300 550 750 height=10. cm,width=14.cm]
{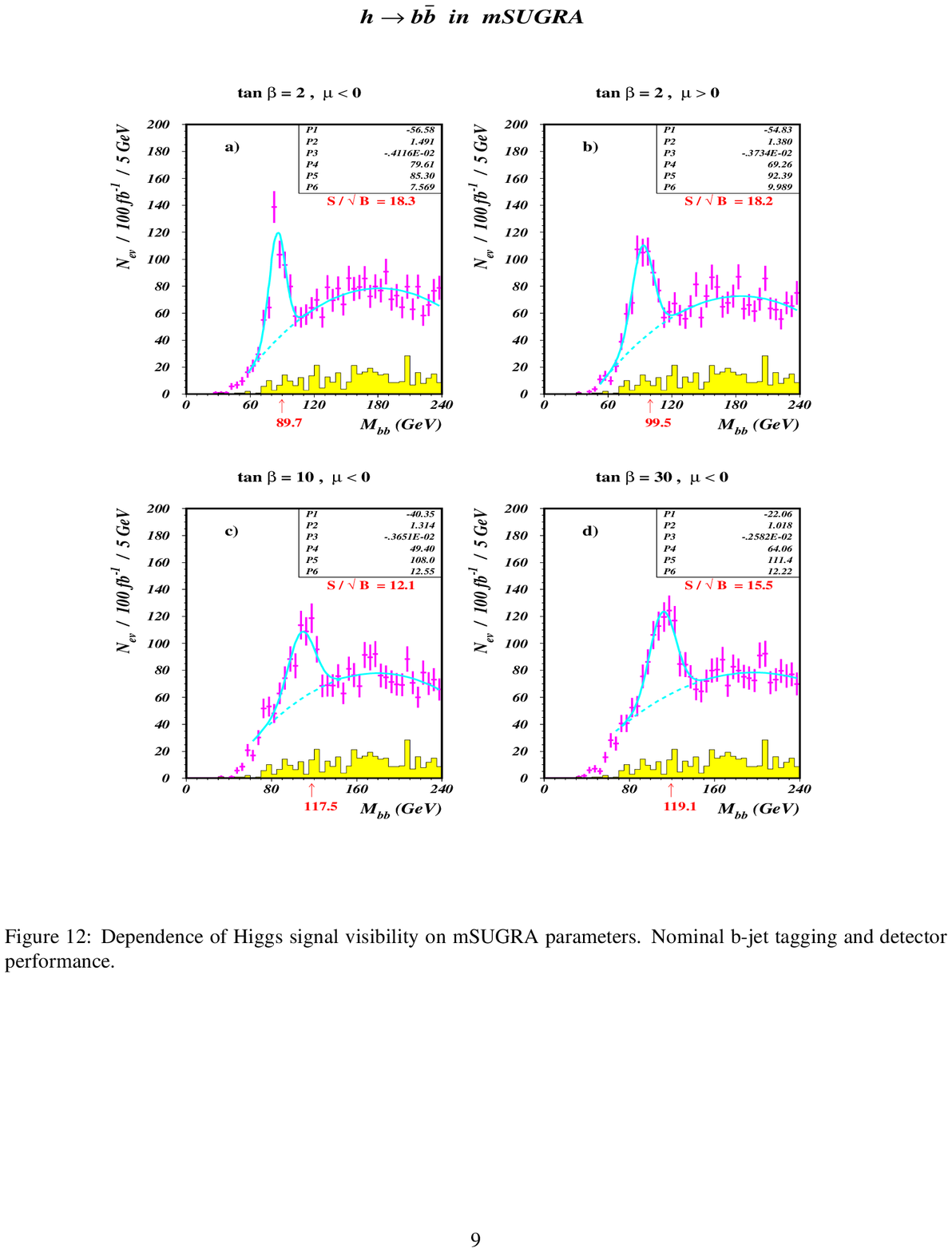}
\caption[fig16]
{CMS simulation of a $h \rightarrow b \bar{b}$ search 
in events with large missing transverse energy 
(from squark and gluino cascade decays) and a luminosity 
of L=100 fb$^{-1}$~\cite{cms98006}. Clear mass peaks are seen for 
various choices of $\tan \beta$ and $\mu$.}
\end{center}
\end{figure}
 
An interesting approach to determine a 
SUSY mass scale has been suggested in a recent ATLAS study~\cite{atlas107}.
The idea is to define an effective transverse event mass,  
using the scalar $p_{t}$ sum of the jets with the largest transverse energy
plus the missing transverse energy of the event. 
One finds that this effective mass shows a reasonable linear 
relation to an underlying SUSY mass, defined 
as the minimum of the squark or the gluino mass.
While this idea appears to be very attractive within the mSUGRA
model, the validity of the proposed relation in 
more general SUSY models remains to be demonstrated.

In addition to the above SUSY studies, 
one would like to get answers to questions like:
\begin{itemize}
\item
What are the branching ratios of various SUSY particles?
\item
Is the accuracy of the various channels sufficient to determine 
the spin of the new particles?  
\item
Do the data allow to differentiate between specific SUSY models?
\item 
Can one find evidence for CP violation in SUSY decays?
\end{itemize} 
At least some answer might be obtained from future
detailed studies of the various decay chains.
We thus conclude this section with the hope that some 
SUSY enthusiasts will try eventually to answer some of these questions
using the expected performance of the LHC and its planned experiments.

\section {SUSY Discovery Strategies for the LHC ~~~....
putting it all together}

To finish our discussion about 
the various SUSY search and discovery strategies 
at the LHC on has to consider the fact 
that the LHC experiments will not provide any 
physics before the year 2005. 
Consequently, quite some preparation time is left.

The most important aspect for the coming years is the 
confrontation of the assumed detector performance 
with reality. First of all, the   
two huge experiments have to be built according to the 
proposed designs. Any of today's physics case studies has thus to be 
kept realistic and should also reflect the expected experimental 
and theoretical knowledge at day 0. 

For example, LHC studies which show 
a wonderful method on how to discover a SM Higgs boson 
with a mass of 50 GeV, can be considered as a waste of time.
A similar judgement could be applied to some ``hard work'' studies, 
which require unrealistic experiments with non existing systematics. 
We do not follow such simple judgement on ``first studies''
as these studies indicate very often the steps towards a realistic 
strategy. 

\begin{figure}[htb]
\begin{center}
\includegraphics*[scale=0.6,bb=0 100 600 730 height=8. cm,width=12.cm]
{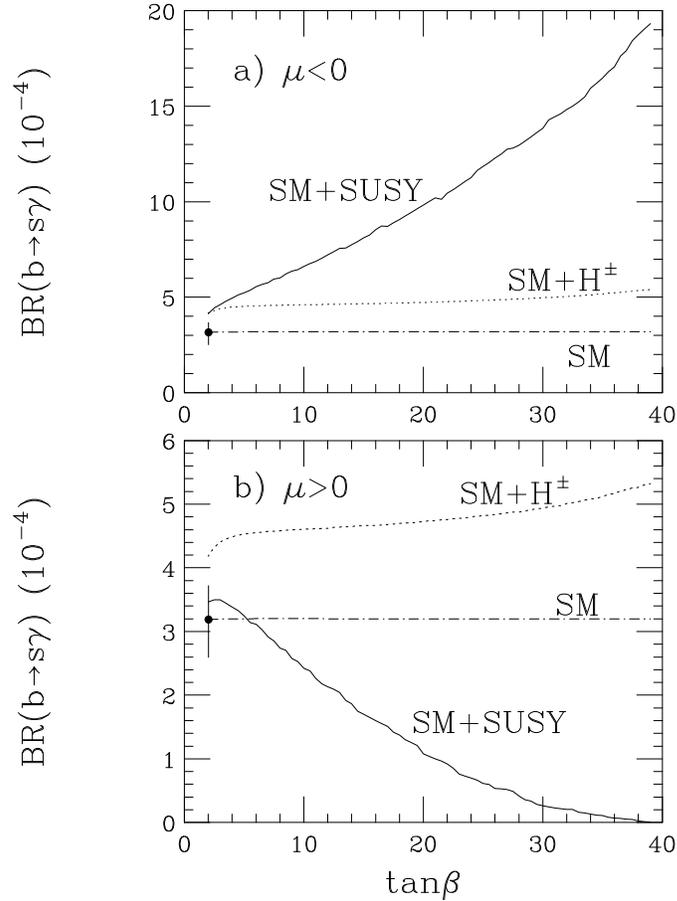}
\caption[fig17]
{Expected branching ratio for $b \rightarrow s \gamma$ 
for the SM and its supersymmetric extension. The  
branching ratio is shown as a function of $\tan \beta$ 
and a negative or positive value of $\mu$~\cite{baer97bsg}.
The preliminary CLEO branching ratio result~\cite{btosgamma} of 
$(3.15 \pm 0.35 (stat.) \pm 0.32 (exp. syst.) \pm 0.26(th)) \times 10^{-4}$
has been added.}
\end{center}
\end{figure}

Realistic and relevant LHC studies should 
be aware of possible constraints from near future experiments.
Examples of such possible constraints come from the LEPII Higgs 
search and the new CLEO result on the branching 
ratio for $b \rightarrow s \gamma$, being 
$(3.15 \pm 0.35 (stat.) \pm 0.32 (exp. syst.) 
\pm 0.26(th)) \times 10^{-4}$~\cite{btosgamma}. 
The near future high luminosity b-factory experiments
should allow to decrease the current error by at least a factor of 4.
  
The current negative Higgs search results from LEP II
exclude essentially the MSSM Higgs sector 
for $\tan \beta \leq 2$.
During the next two years the LEP II sensitivity should increase to  
$\tan \beta$ values of $\leq 4$. Thus, a possible near 
future Higgs discovery at LEPII
requires $\tan \beta$ values between about 2 and 4.
In case that the LEPII experiments will not 
find a Higgs signal, one should study SUSY models with 
$\tan \beta$ values larger than $\approx$ 4.
 
Following some theoretical calculations~\cite{baer97bsg}, the new 
CLEO $b \rightarrow s \gamma$ result, as shown in Figure 17,
appears to exclude a wide mSUGRA parameter range. In particular, one finds
that the mSUGRA parameter $\mu$ has to be positive and that 
values of $\tan \beta > 10$ are essentially inconsistent 
with the existing data. Furthermore,  
future results for the branching ratio
$b \rightarrow s \gamma$ with allowed values between 
$3.5-4\times 10^{-4}$ could exclude the mSUGRA model
well before the start of the LHC. 
On the other hand, in case the near future results require
$b \rightarrow s \gamma$ branching ratios between 
$2-3 \times 10^{-4}$, a 
mSUGRA believer should focus all on  
$\tan \beta$ values between 4 and 10 if  
the Higgs is not seen at LEPII. 
Unfortunately, as can be seen from Figure 18, this $\tan \beta$
range appears to be the most difficult region for MSSM Higgs searches at the 
LHC.
\begin{figure}[htb]
\begin{center}
\includegraphics*[scale=0.55,bb=0 180 600 590 ]
{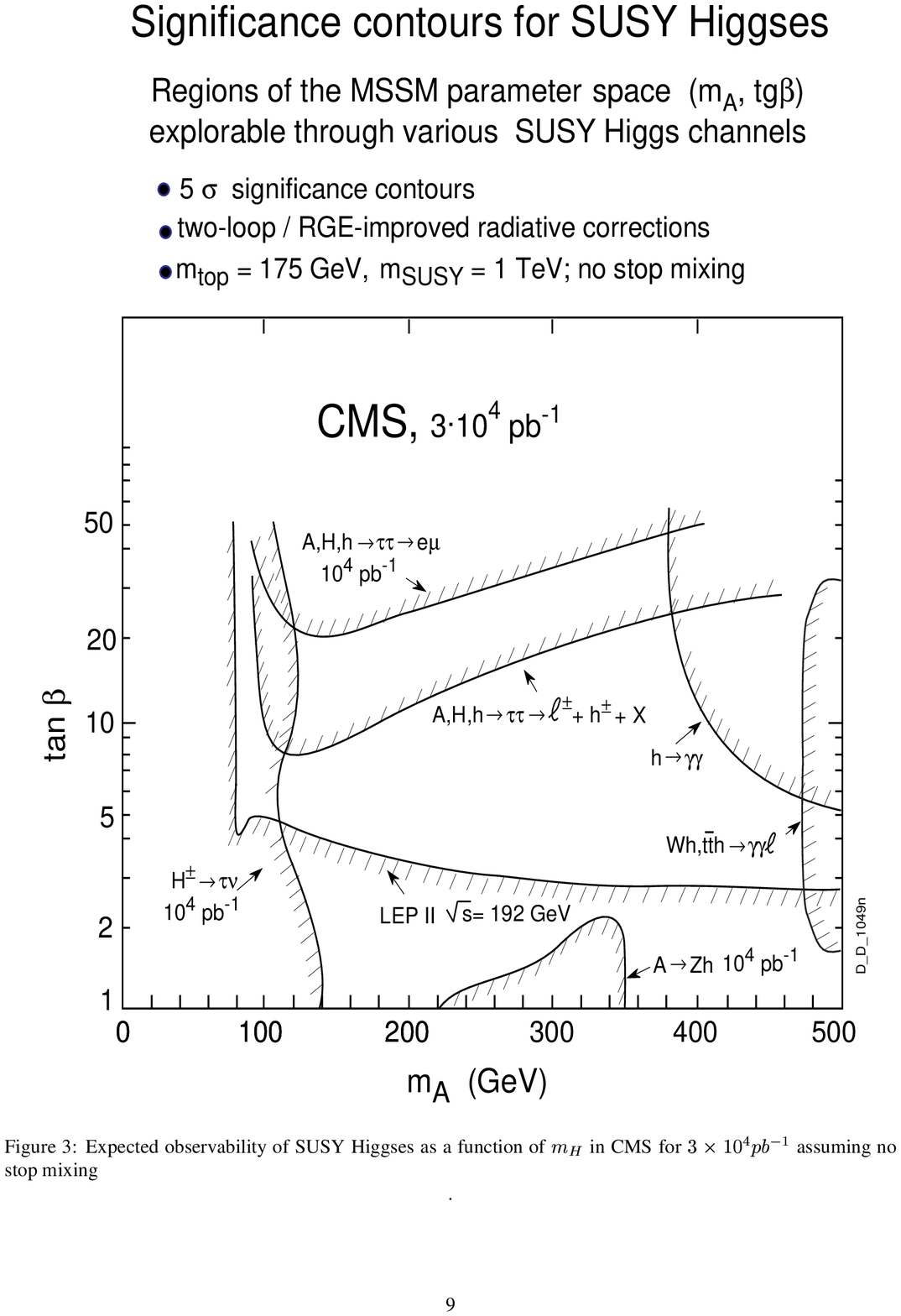}
\caption[fig18]
{CMS 5 sigma significance contour plot for the different MSSM Higgs 
sector in the $M_{A}-- \tan\beta$ plane~\cite{cmsmssmh}.
Each curve indicates the sensitivity for different Higgs search modes.} 
\end{center}
\end{figure}

We thus conclude this section with the remark that 
one should think twice before a too large effort is put into very  
detailed simulation studies as the possible results might be 
proven irrelevant even before such studies are completed!

\section{Super-Summary}

We have discussed a large variety of SUSY particle signatures 
at the LHC. One finds that the ultimate searches for squarks and gluinos
should be sensitive up to masses of about 2 TeV.
However, the large cross sections for 
squarks and gluinos with masses of up to about 600 GeV
should allow to discover SUSY 
during the LHC ``turn on''.

This SUSY discovery potential at the LHC should be confronted 
with the near future expectations from LEPII and the Tevatron.
While only marginal SUSY improvements can be expected, the 
optimistic TeV Run II studies 
expect that the LEPII chargino limit of about 100 GeV 
can be improved with a few fb$^{-1}$ to chargino masses 
of about 130 GeV which might further be increased to values 
as high as 200 GeV at RunIII.
Within the mSUGRA model, one finds that  
the negative searches for charginos at LEPII
imply that gluinos can not be found at RUNII. In contrast, 
negative chargino searches at RUNII and RUNIII 
do still coincide with the ``early bird'' LHC discovery 
potential for squarks and gluinos with masses of about 500-600 GeV.    

Such excellent perspectives
have to be matched however with an almost perfectly working 
full detector and the accurate knowledge of all SM background 
processes. In addition, a well prepared search should consider
a large variety of models and the resulting possible signatures.
 
We would like to finish our review of 
``SUSY Discovery strategies at the LHC'' with a few related quotes:
\newline
\vspace{0.05cm}

{\it ``Experiments within the next 5--10 years will enable us to decide 
whether supersymmetry, as a solution to the naturalness problem of the 
weak interaction is a myth or reality''} H. P. Nilles 1984~\cite{mssm84}  
\newline
\vspace{0.05cm}

{\it ``One shouldn't give up yet'' .... ``perhaps a correct statement is:
it will always take 5-10 years to discover SUSY''} 
H. P. Nilles 1998~\cite{nilles98}
\newline
\vspace{0.05cm}

{\it ``Superstring, Supersymmetry, Superstition''} Unknown
\newline
\vspace{0.05cm}

{\it ``New truth of science begins as heresy, advances to orthodoxy and ends 
as superstition''} T. H. Huxley (1825--1895).
\newline
\vspace{0.05cm}

{\bf \large Acknowledgements}
{\small I would like to thank the organisers of
the SUMMER school on ``Hidden Symmetries and Higgs Phenomena'' in ZUOZ
for the invitation to review SUSY search strategies at the LHC.
The interesting discussions about hidden phenomena were 
certainly also stimulated by the alpin landscape with 
many beautiful peaks. I am grateful to D. Haztifotiadou and Z. W\c{a}s 
for their critical suggestions and comments about my lectures and 
its writeup.}

\end{document}